\documentclass{iaus}
\usepackage{graphicx}
\usepackage{amsbsy}
\usepackage{AMSsymb}
\newcommand{\turck}{\textrm{Turck-Chi\`eze }}

\title[IAU 264.~Solar and Stellar Activities: Impact on Earth and Planets] 
{Concluding remarks on Solar and Stellar Activities and related planets}

\author[Sylvaine Turck-Chi\`eze]   
{Sylvaine Turck-Chi\`eze}

\affiliation{SAp/IRFU/CEA CE Saclay, 91191 Gif sur Yvette cedex, FRANCE \\email: {\tt sylvaine.turck-chieze@cea.fr}}

\pubyear{2009}
\volume{264}  
\pagerange{1--126}
\jname{Solar and Stellar Activities: Impact on Earth and Planets}
\editors{A. H. Andrei, A.S. Kosovichev \& J. P. Rozelot, eds.}
\begin{document}

\maketitle

\begin{abstract}
The symposium has shown the dynamism of this rapidly evolving discipline. I shall 
concentrate  here on some highlights and some complementary informations.  I conclude on open questions with some perspectives on solar \& stellar activity and related planets.   
\end{abstract}

\firstsection 
\section{Introduction}

This symposium was extremely exciting,  thanks to the interaction between several communities working on the Sun, star formation, young stars and X-rays, the Earth atmosphere and XUV impact on planets and life. Consequently the coverage area of the field is very broad from the Sun to the Earth coupled to this emergent and very active new field dedicated to the interactions between young Stars and Planets. We have all noticed that the novelties of the last  ten years were very important.  This field has developed very rapidly and the symposium was extremely rich in new results, with 50 oral presentations and 110 posters.  Of course the idea of this paper is not  to do a summary of the previous papers but to emphasize highlights with useful complementary informations and list  some open questions and perspectives which can be deduced from the present situation.

We are clearly in a lucky period because through all the interactions and shared interest,   a new community is emerging with common questions and common physics. The situation was radically different a decade ago. There was rather  few studies on the connection between the stellar interior and the solar emergent flux,  chromosphere and corona. Today, the magnetic field is a link between these communities. The Sun (Star) and Earth (Planets) connection is another example. This field has emerged at the preceding IAU but we see in this symposium that communities are now well organized to discuss this(ese) field(s) in its complexity, with potential  consequences for life on planets.

We are lucky also because the symposium appears during a special year where the solar minimum is particularly unusual  with an exceptionally long absence of  sunspots,
in addition to  a very long natural eclipse and a lot of satellites in operation. This scientific fact may guide science slightly  differently and this is also a very good thing.

Finally we are lucky because we get more and more observational evidences of variability and we are waiting new  space and ground based instruments, so it is a good moment  to determine what we have already understood, what are the questions on which  we may hope new  progress,  what are the questions for the next generation of instruments and what kind of instruments do we need. 
 
 I recall here some highlights and add complementary informations. I shall present also a list of open questions and some perspective. I shall separate this review in three parts:
 
 - the solar (stellar) fundamental quantities and their variability
 
 - the solar internal variability, its emergence and other stellar variability
 
 - the Stars and Planets interaction (s).
 
\section{The solar (stellar) fundamental quantities and their variability} 

The solar fundamental quantities are the total irradiance, the radius and the composition. Table 1 summarizes their values and their variability with the 11 year solar cycle. It is quite amazing that the calculation of a solar standard model requires the  adjustment of the initial helium, the initial metal composition and the mixing length to get the measured luminosity, radius and surface abundances at an accuracy level of 10$^{-4} $- 10$^{-5} $ when the star of 1 M$_\odot$ star reaches the age of 4.6 Gyrs (including the contraction phase and the pre mainsequence). One notes that this accuracy  is not yet reached today.  In the standard framework, the solar constant does not vary by more than 10$^{-8} $ during the last 100 years.  Of course this is contradicted  by the irradiance measurements.

\subsection{The solar irradiance} 
Clearly the Sun does not agree with the standard model: its luminosity varies by less than 10$^{-3}$ every 11 years, the Hale cycle,  about 0.9 W/m$^2$, but in fact 3 or 4 times more near the maximum of the cycle due to the presence of sunspots and faculae (Figure 1). These last two years, the total irradiance has been measured at a real minimum, characterized by the absence of sunspot:  a decrease of about 0.25-0.3 W/m$^2$ in  comparison with previous cycles 21 and 22 minima has been obtained. But, if there is a  general trend on the long term, it is not yet established. Of course the interpretation of the cyclic variability of the solar irradiance is attributed for a large fraction (at least 80\%) to 
the radiative effects of the magnetic activity in the photosphere (\cite{Domingoetal09}). 

{\it Does the total irradiance variability reflect only  superficial  magnetic effects and the Hale cycle ? If there are some other sources, what are they ?}

\begin{table}
  \begin{center} 
    \caption{Evolution of the solar fundamental constants and their variability along the 11 yr solar cycle. The heavy elements fraction mass comes from Anders \& Grevesse estimate, Grevesse \& Noel one and today from Asplund et al. (2009). See text.}
    \begin{tabular}[h]{lcccc}
     \hline
     & Reference values Allen &  Present values &Time Variability  \\
 \hline
Luminosity  & 1360.488 ($\pm 2. 10^{-4}$)  &   1367.6 W/m$^2$ - 1361    W/m$^2$  &  1-4  W/m$^2$  \\
Radius   &   695 990 km  &   693710 (min)   &     10-160 km \\  
Seismic radius (f modes)&        -       &               695660 km             &          -            \\
Radius shape  &   -     &   oblateness 6 to 10 km    & 6-14 km  \\
Heavy element Z & 0.02 then 0.0173 &  0. 0134  &  no evidence\\
in fraction mass  &                &                    &\\  
  \hline
\end{tabular}
 \label{tab:fund}
    \end{center}
\end{table}
\normalsize
\begin{figure}
\begin{center}
\includegraphics[width=7cm]{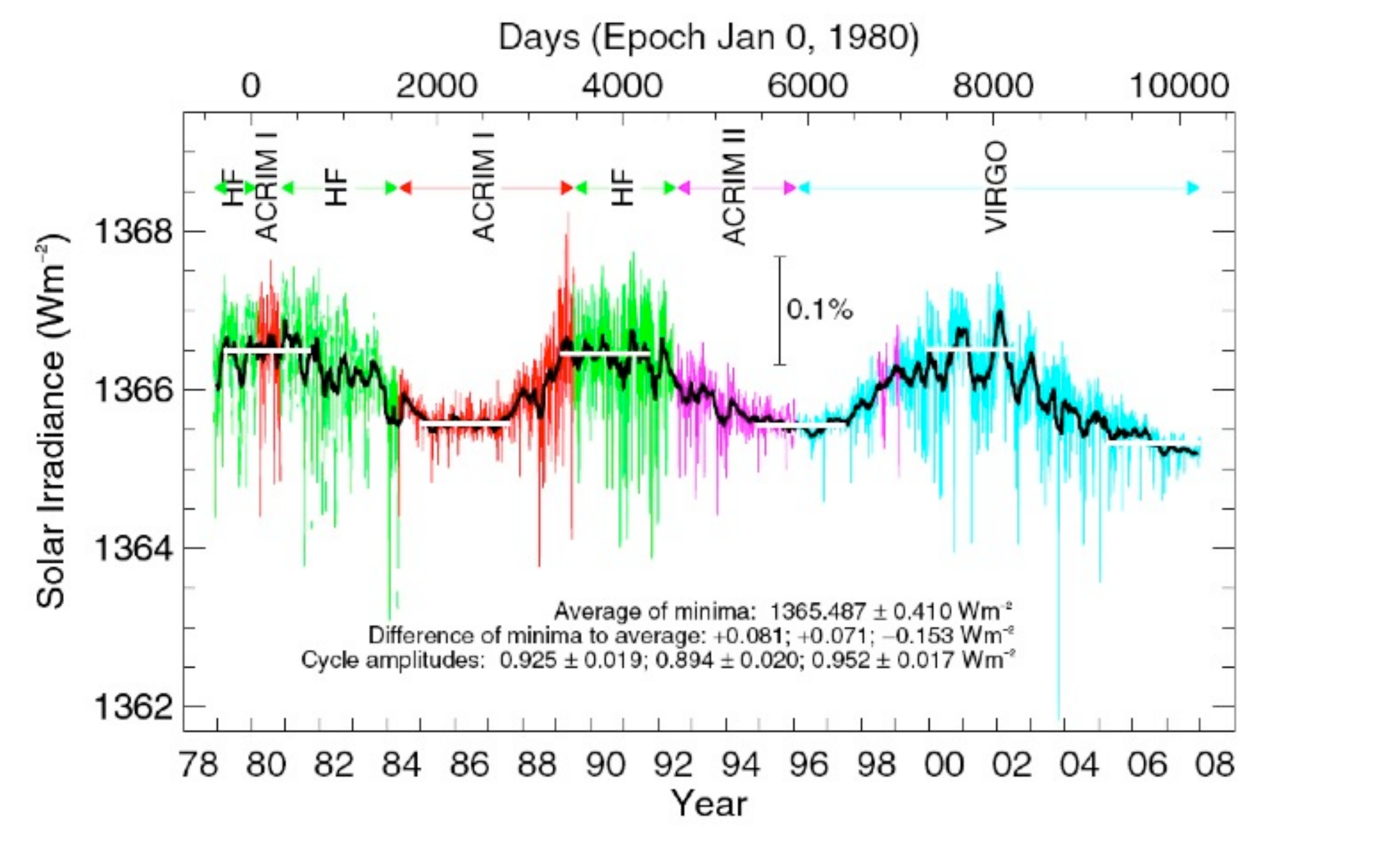}
\includegraphics[width=6.cm]{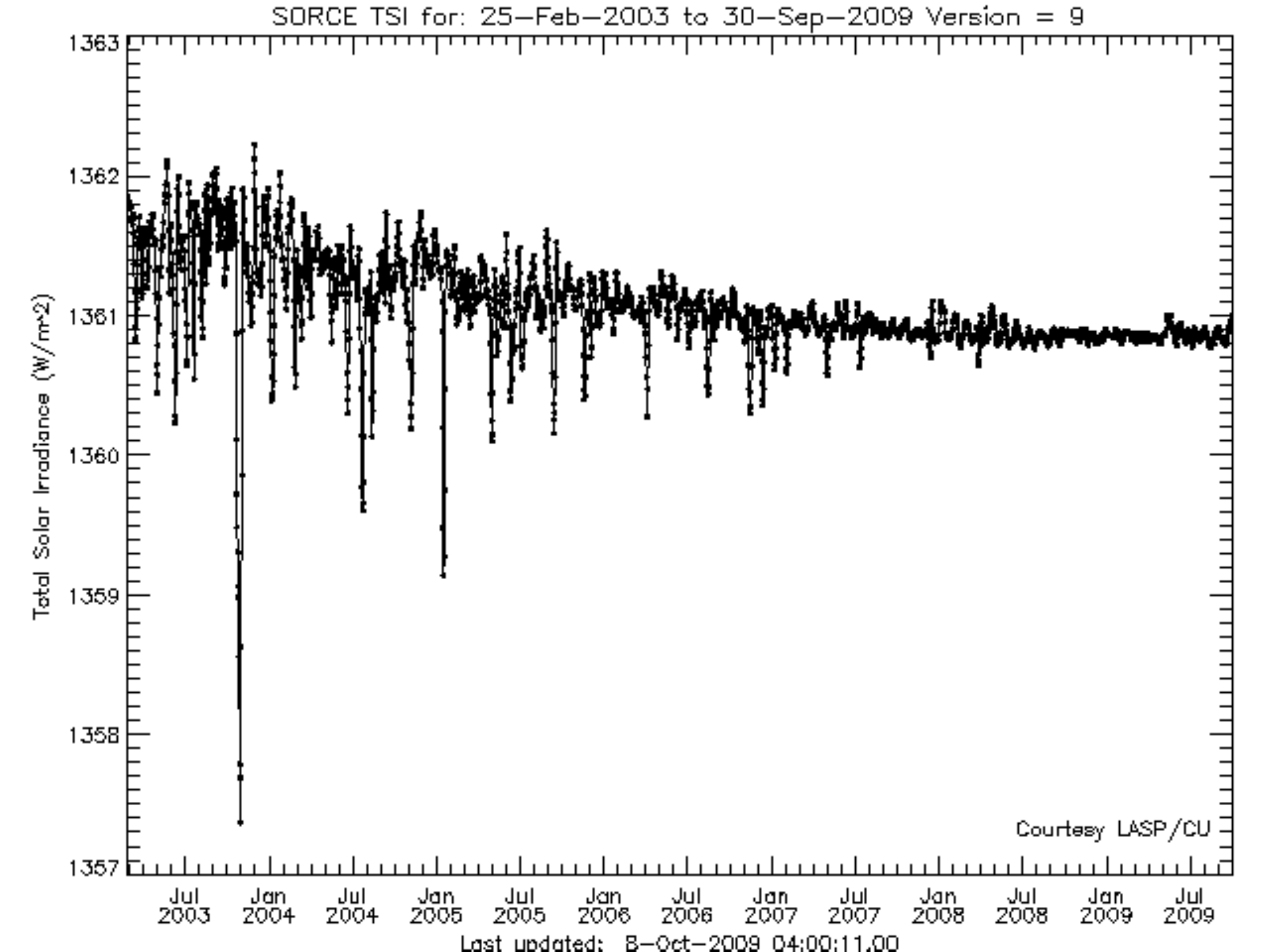}
\caption{a) Reconstruction of the solar irradiance variability obtained by ACRIM and SoHO satellite observations. From Fr{\"o}hlich (2006). b) Zoom on  the TSI  measured by SORCE from July 2003 to September 2009 \cite{Kopp05}. One notices a rapid strong variability due to spots and faculae on both datasets near the maxima. The absolute value difference between the two series has no explanation up to now.}
\label{fig:tim}
\end{center}
\end{figure}
\begin{figure}[h]
\begin{center}
\includegraphics[width=16.5pc]{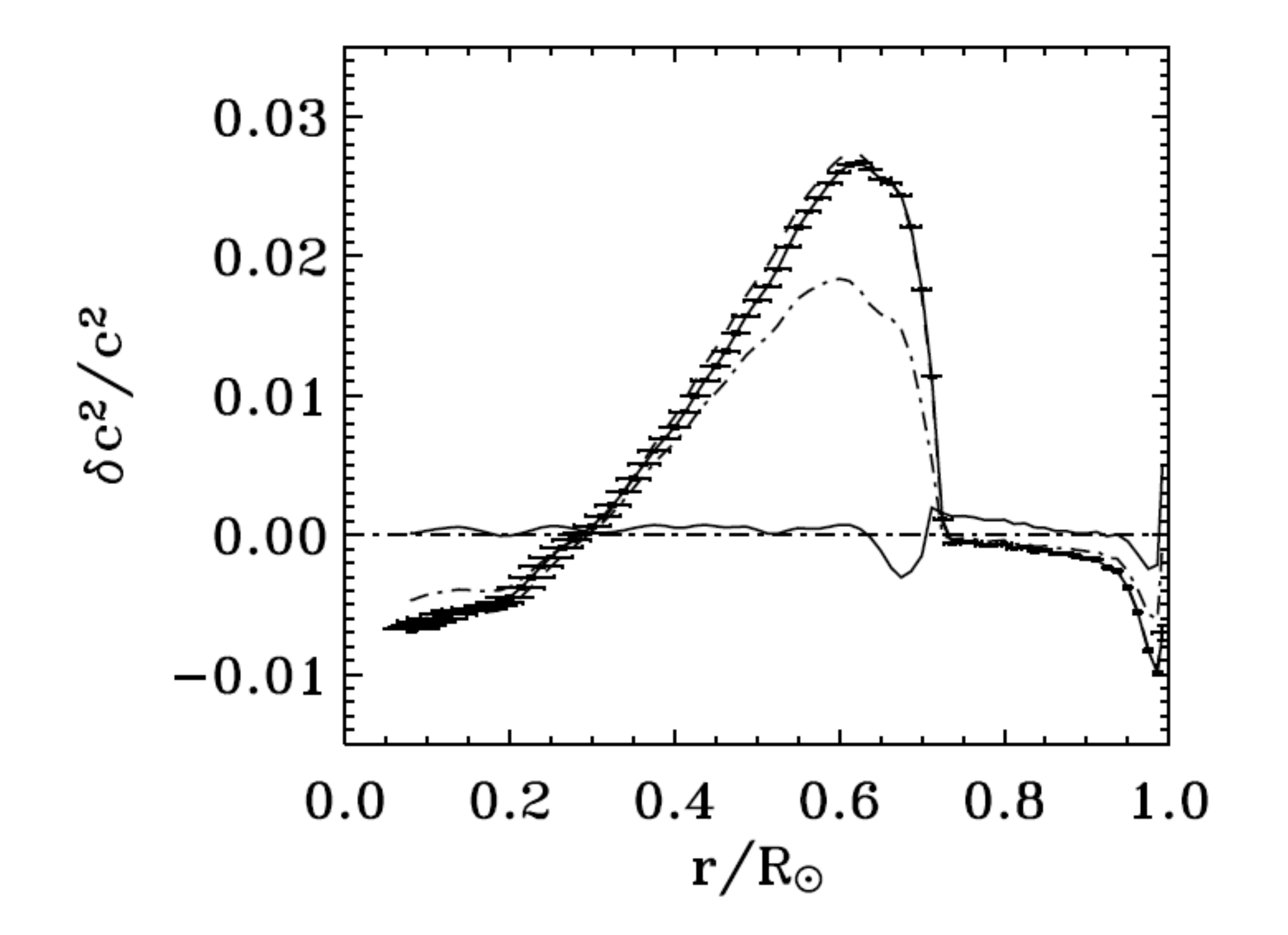}
\includegraphics[width=15pc]{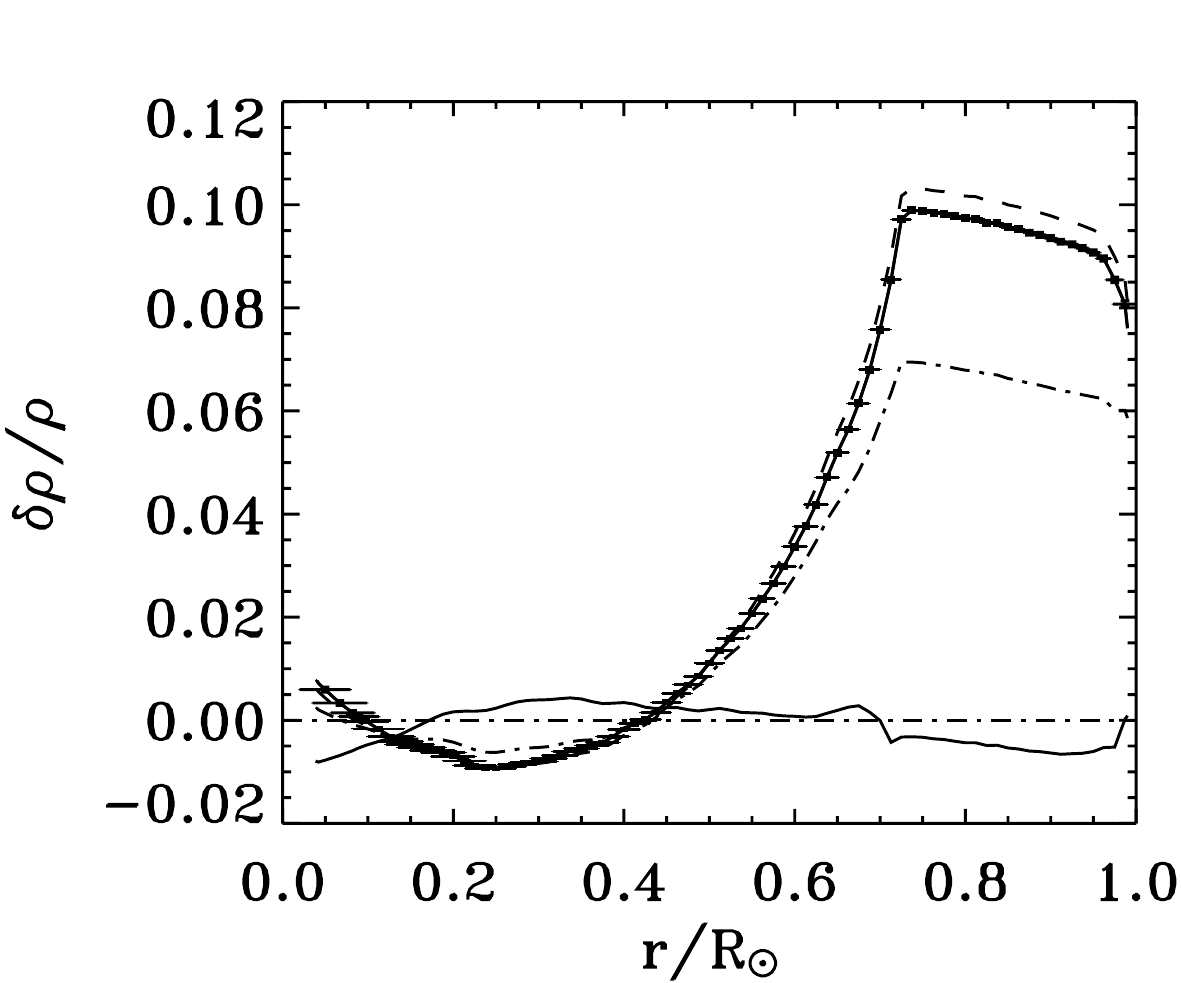}
\caption{ \label{label} \small Squared sound speed and density differences extracted from the seismic inversions obtained with GOLF+MDI/SOHO acoustic modes and the standard model 
using the Asplund 2005 composition (solid line with seismic error bars) or using Holweger (2001) composition (dot  dashed line) or Lodders (2003) one (dotted line). The difference with the
seismic model is drawn in full line. The vertical error bars are too small  to be visible, the horizontal ones are still large in the nuclear core (below 0.3 $R_\odot$ which contains more than half the mass of the Sun), they will be strongly reduced in adding several gravity mode frequencies. The present composition is compatible with such differences. From \cite[\turck et al. (2004a)]{Turcketal04a}.\normalsize}
\end{center}
\end{figure}
\subsection{The detailed solar composition} 
The solar abundances are measured through the photospheric absorption lines. Their determination has been improved several times and all  suspicious problems (photospheric iron in disagreement with meteoric one, absence of comparison with meteoric elements for C, N, O, Ne) seem to be solved. Today, two teams are converging toward an abundance of heavy elements (Z= 0.0134) strongly reduced in comparison with what was used 20 years ago  (\cite{Asplundetal09}; \cite{Caffau09}). This reduction is in fact a little bit smaller that announced 5 years ago and in reasonable agreement with Holweger estimate \cite[(Holweger et al., 2001)]{Holwegeretal01}.  This is an important result because of its consequences on galactic evolution: the Sun is no more considered as enriched by a supernova \cite[(\turck et al., 2004a)]{Turcketal04a}. This recent result confirms the inability of the standard model to predict properly the observed sound speed (see Figure 2). One now needs to look for the processes which justify this disagreement: microscopic diffusion, opacities, dynamical proocesses (\cite[\turck et al., 2004a]{Turcketal04a}; \cite[\turck et al., 2009a)]{Turck_etal09}. 

{\it Up to now, there is no reason to seek an explanation based on microscopic processes rather than  on macroscopic processes, both must be investigated (see below)}.

\subsection{The solar radius} 
The time variability of the radius and the photospheric shape are  rather difficult subjects because the effects are very small. Progress on this field is important these last years and several issues have been clarified: in using radiative transfer calculations, \cite[Habbereiter et al. (2008)]{Habbereiteretal08} estimate that the solar radius measured at the limb differs from the  f-modes estimate by 0.3 Mm (+ 4.3 10$^{-4}$). It is not so easy to estimate the time  variation of the radius with the solar Hale cycle but a detailed analysis seems to show an anticorrelation (Lefebvre \& Kosovichev, 2005; Lefebvre, Nghiem \& Turck-Chi\`eze, 2009). This anticorrelation is  also suggested by ground and balloon measurements. But the space and balloon variability differ by one order of magnitude (between 10 to 160 km) along the solar cycle but  they use different techniques at  different wavelengths \cite[(Egidi et al., 2006]{Egidietal06}; \cite[Emilio et al., 2007)] {Emilioetal07}. 
The MDI data point to nearly purely oblate shape near
solar maximum, but with a significant hexadecapole shape near minimum. Of course it would be nice to follow the time radius variability on ground, with a well known technique as an astrolabe. The brazilian team has shown the difficulty due to atmospheric effects but the interest of continuity at low cost encourages to pursue ground measurements. They get a mean value of the radius of  959.163 $\pm$ 0.004 arc. \cite[(Andrei et al., 2006)]{Andrei06}, which can be compared to the solar minimum value obtained with the SDS balloon of 959.561$\pm0.111$  \cite[(Djafer, Thuillier \& Sofia, 2009)]{Djafer08}. The aspheric shape of the Sun is not yet understood, the balloon shows values of oblateness of 4.3 to 10. 10$^{-6}$ increasing with the cycle \cite{Emilioetal07}. This can be compared to the effect of the internal rotation alone which leads  to values of the order of 2.-3. 10$^{-7}$. It is evident that the role of the deep and varying magnetic field must be added. See a complete review on the subject from \cite{Rozelot09}.

{ \it The present situation seems today unclear. The methods used to extract the information must be compared, and very precise measurements from space might clean the present situation}.
\subsection{The other stars}
The situation is largely different for the other stars. In most cases, the radius is not known, only its luminosity and effective temperature are determined. Consequently the mass is not precisely estimated, and depending on the detailed physical ingredients considered (composition, overshooting...) the mass can vary up to a factor 2. The radius is only measured for well deformed stars which rotate quickly. We have very good hope that the deployment of asteroseismology with COROT and KEPLER will largely improve the situation, the asteroseimic revolution arrives clearly with KEPLER which delivers observations of thousand stars from the Lagrangian L2 point. The main  interest related to this conference is to better establish which kind of activity is found in  different stars.
 
\section{Solar variability and other stellar types of variability} 
It is now relatively well established that the solar and stellar activity originates from the internal processes, but it is only recently that quantitative works have been devoted to the link between internal and external phenomena.  We note presently very important specific results but not yet a general picture. Once again the Sun remains the best guide to study this connection in details.
\subsection{The Sun is highly observed globally} 
We are lucky because the ground based instruments are now accompanied by space missions like Ulysses, SoHO, RHESSI and now STEREO and HINODE. Altogether, they offer a lot of complementary information on the solar external manifestation of activity. The variability in sunspot, $\gamma$, X and radio  will be treated in the next section. I shall concentrate here on global effects. Figure 3 summarizes the evolution of ideas in the understanding of the solar magnetic field with the Ulysse spacecraft. Prior to Ulysses, the models of the solar magnetic field assumed  that it was largely dipolar, field lines near the solar equator were thought to form closed loops whereas field lines from the poles were dragged far into interplanetary space by the solar wind (Figure 3 left). Ulysses (Figure 3 right) found that the amount of outward magnetic flux in the solar wind did not vary greatly with latitude (\cite {Smith03}).  Nevertheless,  the solar wind has been observed during three extrema and it is now measured daily aboard SoHO. The slow and high velocity components  are not always present simultaneously, the high component depends on the importance of  the coronal holes from where emerges the fast wind. Figure 4 illustrates how different are  the two last minima. The present one is very long and is characterized by the absence of sunspots. It is accompanied by a low latitudinal wider wind and  a less quadrupolar configuration of the wind.  Its global pressure is decreased by 20\% and the magnetic field at the level of the spacecraft  reduced by 36\% (\cite{McComas08}). These results are of prime importance for the extension of the heliosphere which is presently reduced and consequently  inhibits more the protection against the cosmic rays. This information is extremely useful for Space Weather (see below).
\begin{figure}
\begin{center}
\includegraphics[width=5.3cm]{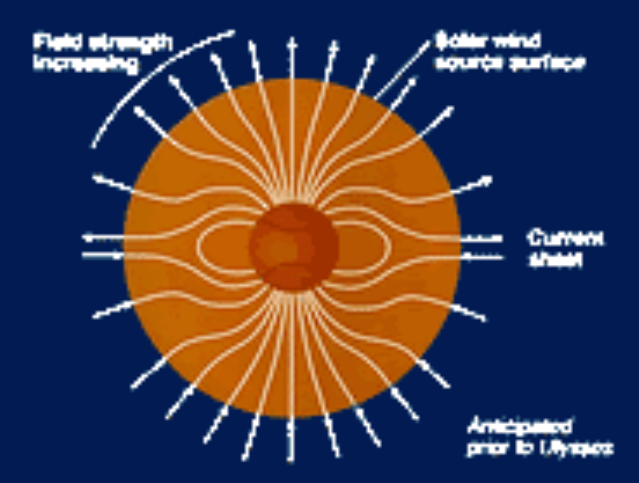}
\includegraphics[width=6.cm]{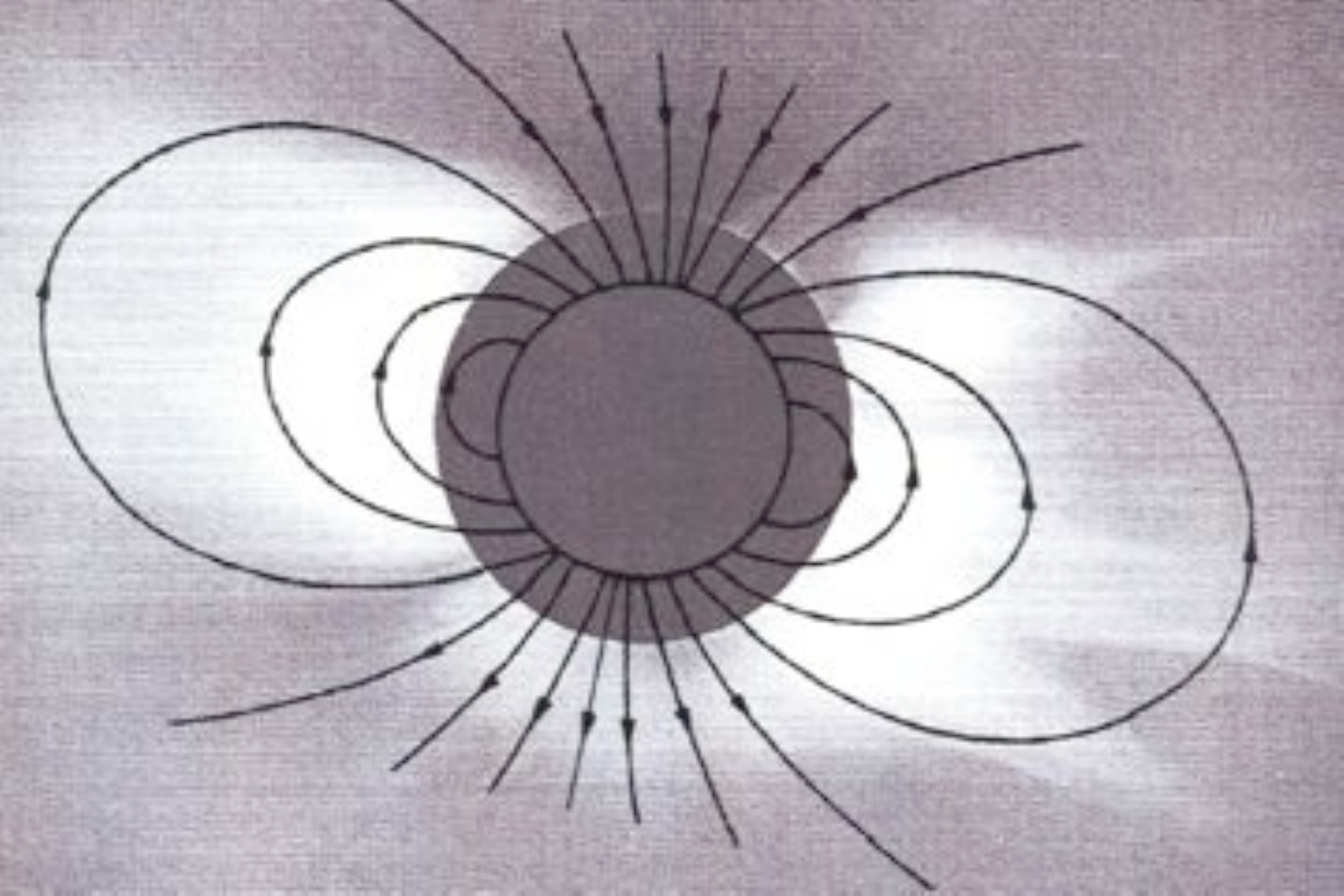}
\caption{Evolution of the ideas on the configurations of  the solar magnetic field  thanks to the 18 years of ULYSSES spacecraft  observations  out of the ecliptic. }
\label{fig:conf}
\end{center}
\end{figure}
\begin{figure}
\begin{center}
\includegraphics[width=12.cm]{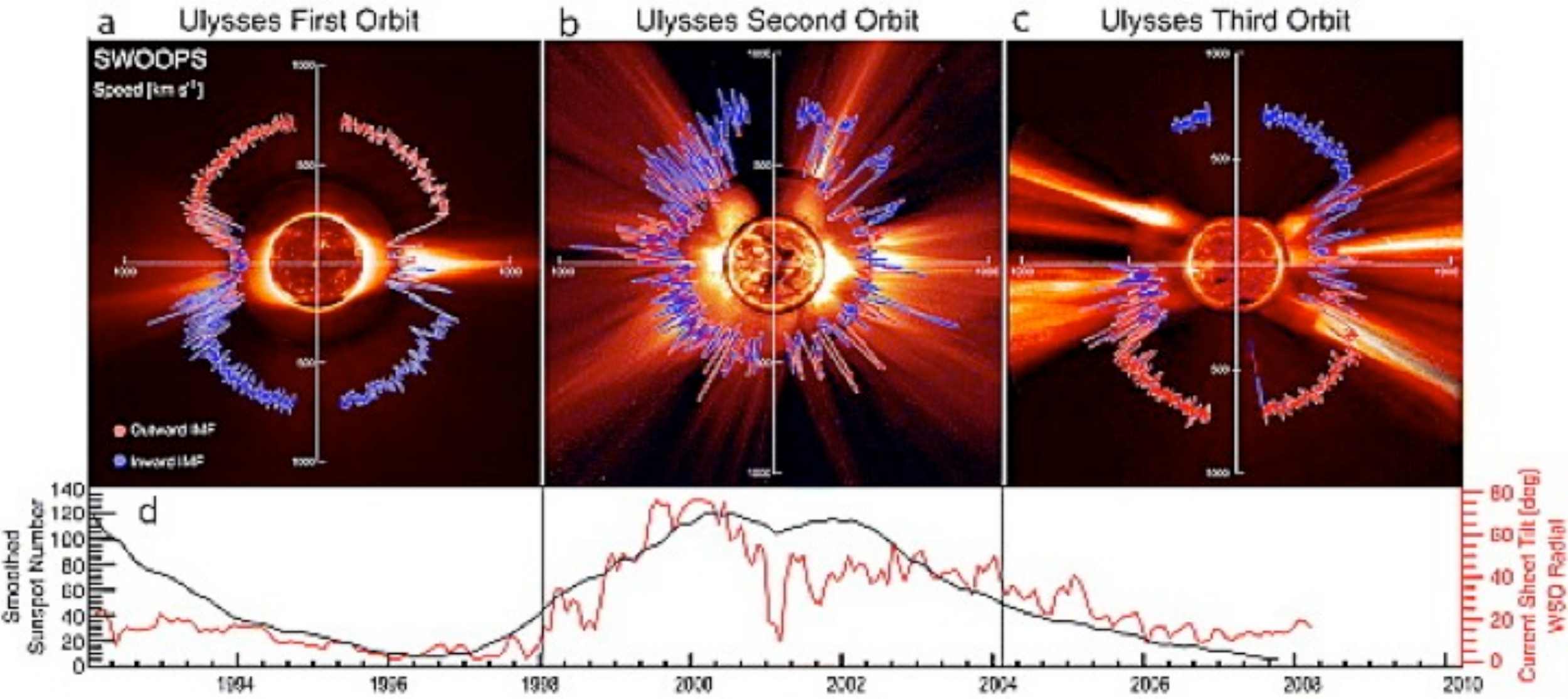}
\caption{Polar plots  of the solar wind seen by Ulysses during two successive minima a), c) and one maximum (b). Superimposed are images from EIT/SoHO. d) time variation of the smoothed sunspot number and of  the heliospheric current sheet tilt. From McComas et al. (2008).}
\label{fig:conf}
\end{center}
\end{figure}

 In parallel, W. Schmutz mentioned at the 3rd Climate Symposium in Lapland (see its oral presentation on website) that the interplanetar magnetic field  B$\rm_{IMF}$ is a more justified indicator of activity than the solar sunspots which are proxies  for active regions.  The total irradiance TSI is well correlated with B$\rm_{IMF}$ which  takes into account the quite Sun, network and active regions. He then deduces a value of 2 nT for B$\rm_{IMF}$ and a TSI of 1364.4 W/m$^2$ at the Maunder minimum, by extrapolation of the three decade observations of TSI  obtained aboard the ACRIM and SoHO missions (\cite{Frohlich06}). 
\begin{figure}
\begin{center}
\includegraphics[width=10cm]{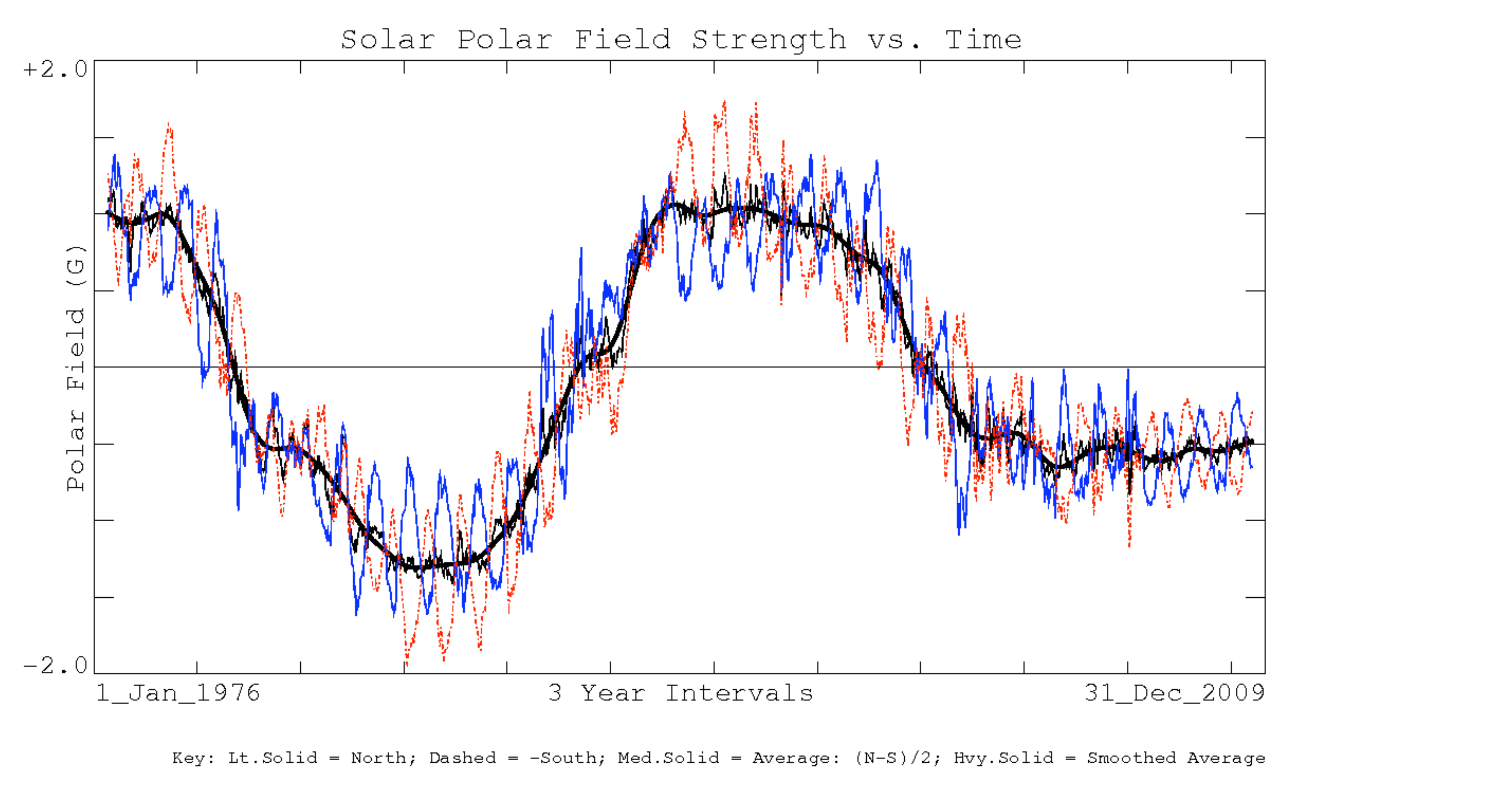}
\caption{Solar polar field for latitude greater than 65 degrees measured at the Wilcox Observatory along the last 30 years. One observes a very different trend of this quantity between the two successive minima. See Hoeksema talk and Schatten (2005).}
\label{fig:conf}
\end{center}
\end{figure}
\begin{figure}
\begin{center}
\includegraphics[width=7.cm]{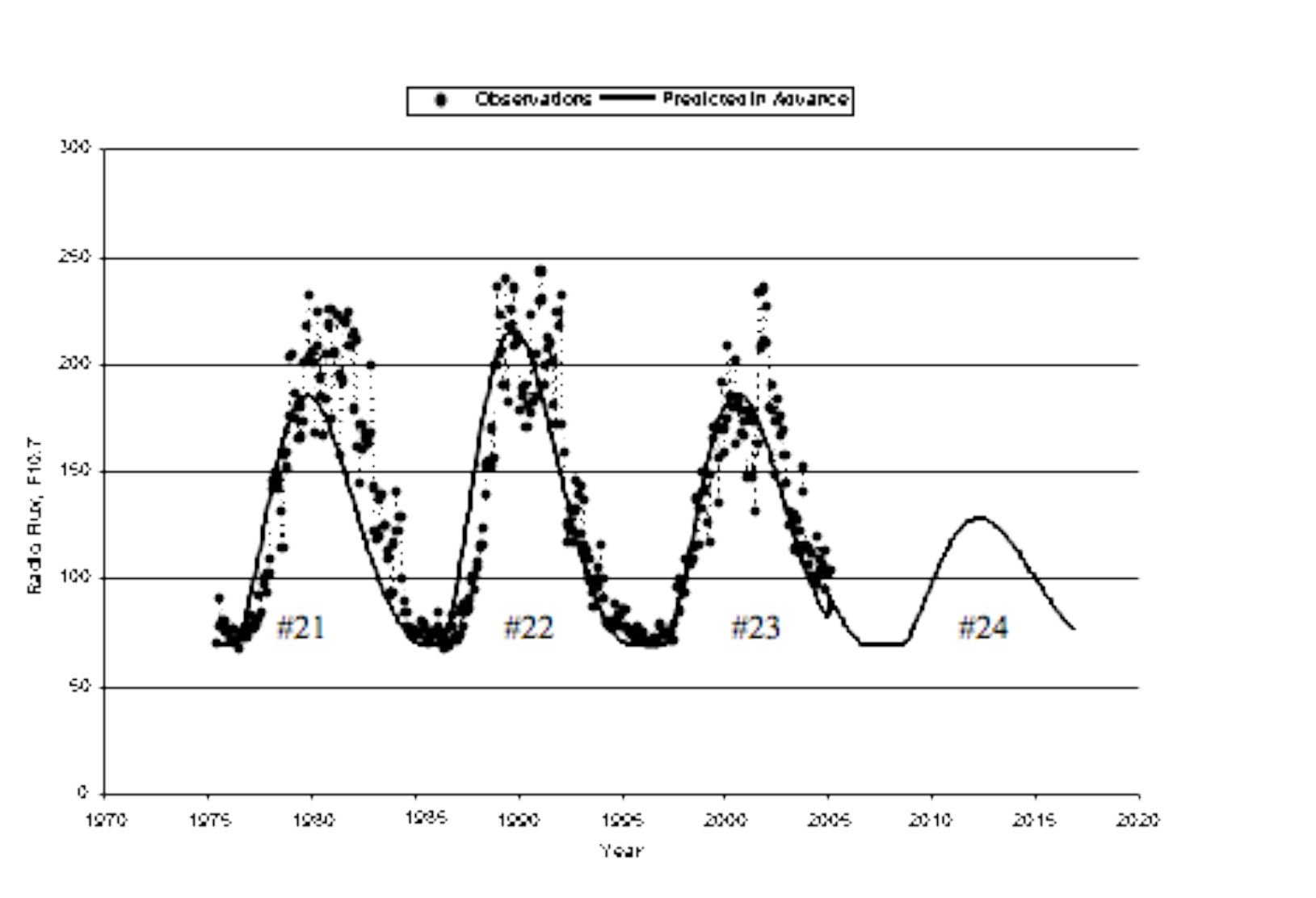}
\includegraphics[width=6.cm]{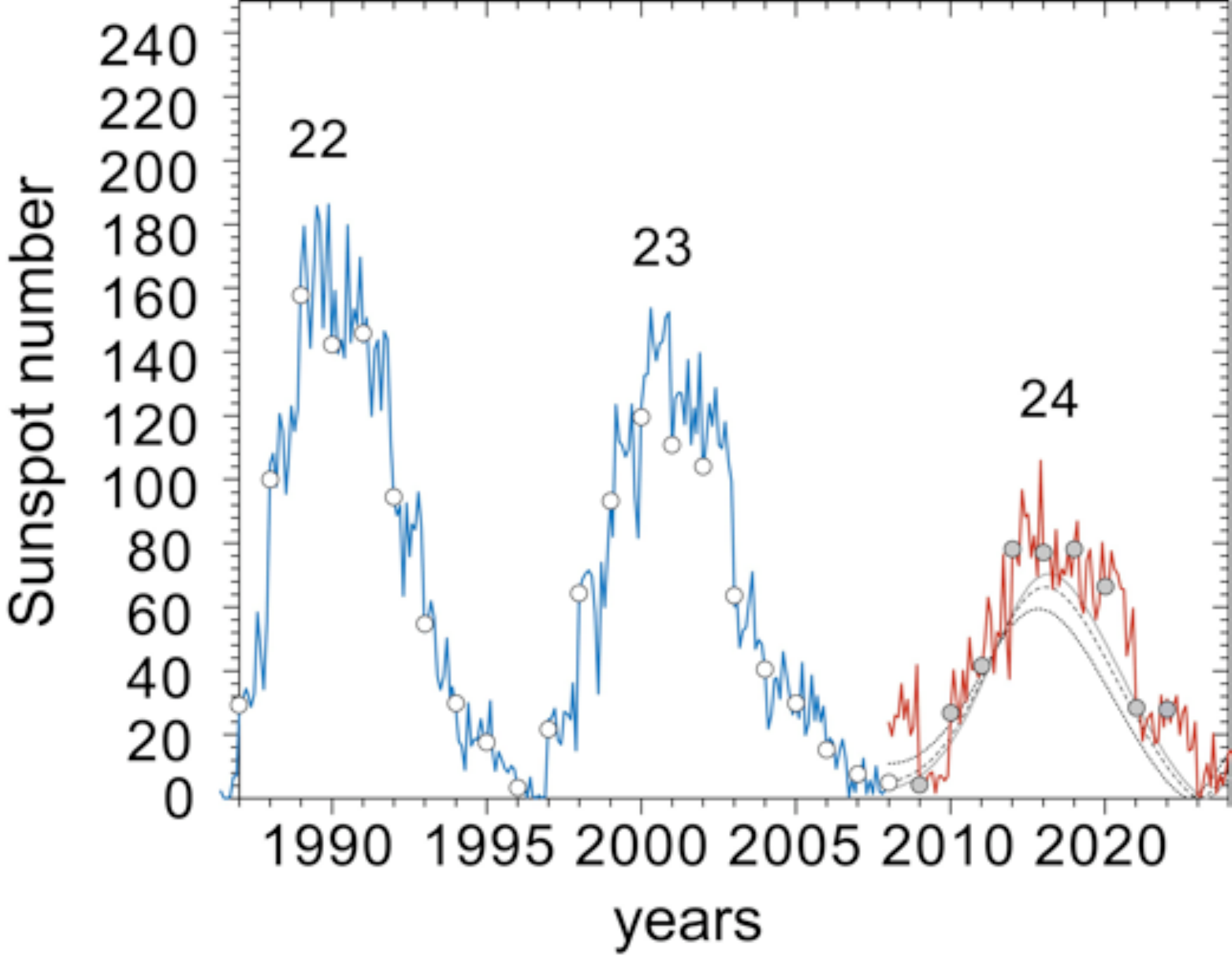}
\caption{ Observed F10.7 Radio Flux (circles) and solar flux predictions (solid lines) prior to each of the last three cycles. The radio flux is in 10$\rm^{-22} J s^{-1} m^{-2} Hz^{-1}$. From Schatten (2009). Assimilation of data to predict cycle 24 done by \cite{Kitiashvili08}.}
\label{fig:cycle}
\end{center}
\end{figure}
Nevertheless, the prediction of the time evolution of the Hale cycle remains difficult. There has been very few predictions of the present very long low activity period. But a lot of indicators have shown{\it a posteriori} that  2004 has been an informative year. It is directly comparable to the previous minimum (cycle 22). At the Wilcox Solar observatory, 30 years of data have been accumulated on the antisymmetric and symmetric zonal flows. It is the same for  the polar field (above 55 degrees) which varies between +1 and -1.5 G (see Figure 5) and the daily solar mean magnetic field which varies between 1 to -1 G.  Now MDI and HINODE give also access to the polar field above 75 degrees and confirm a different behavior for the two cycles. Cycle 23 keeps a high value of about 7 G during at least 4 years in contrast with cycle 22. Zonal flows below the surface show also slower velocities than during the previous minimum \cite{Howe09}.

{\it What is (are)  the origin(s) of the polar field ? Fossil or Convective dynamo field ?}  
\subsection{The solar variability and the internal dynamical processes} 
\subsubsection{The solar Hale cycle and the convective zone} 
\begin{figure}
\begin{center}
\includegraphics[width=10.cm]{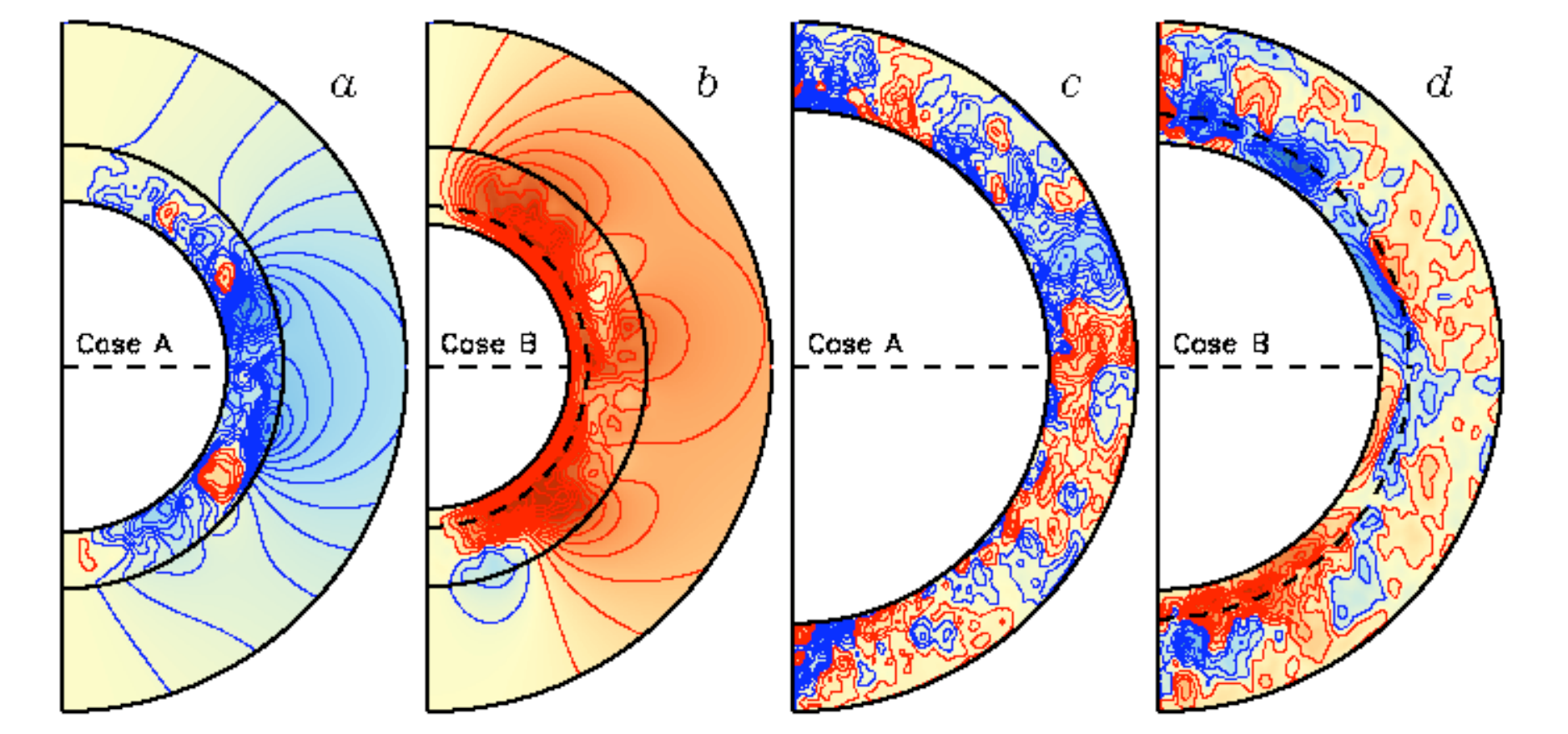}
\caption{Mean solar convective poloidal magnetic fields (a, b) and  toroidal magnetic 
fields(c, d) for two simulations differing by their magnetic Prandt numbrr (respectively 4 and 8), averaged over longitude and time 
(90 days). White (red in online version) tones denote a clockwise poloidal 
field orientation and eastward toroidal field whereas black (blue) tones de- 
note the opposite. Saturation levels for the poloidal and toroidal color ta- 
bles are ±20G and ±2000G respectively. Poloidal fields in (a, b) include a 
potential-field extrapolation out to 1.5R and the base of the convection zone 
is indicated by a dashed line in (b, d). From Miesch et al. (2008).}
\label{fig:MF}
\end{center}
\end{figure}

It is fundamental to understand the solar variability. The task is difficult because we have no direct access to the topology and strength of the internal fields.  Important progress has been obtained during the last decade on the understanding of the Hale dynamo originally based on the works of \cite{Babcock61} and \cite{Parker84}.   It is interesting to notice that a shallow dynamo has a good predictive character for the next cycle. Effectively in the Babcock-Leighton Flux Dynamo theory, the polar field at the end of a cycle serves as a seed for the next cycle's growth. By examining 8 previous solar cycles and their impact on Earth, the method has successfully predicted cycles 21, 22 and 23 and 24 (Figure \ref{fig:cycle} left) using WSO and MWSO data. The unusually weak polar field around 2005 leads to forecast a small cycle 24  (\cite{Schatten03}; \cite{Schatten05}; \cite{Schatten09}). A more fundamental 2D approach explores  a large scale dynamo that implies the whole convective zone and the tachocline. They solve the induction equation and consider not only the $\alpha$ and $\Omega$ effects but also the meridional circulation in the whole convective zone together with an estimate of the velocity on at least the last 3 solar cycles. They try to understand how the field is regenerated (\cite{Dikpati08}).  Such prediction is improving with time and the  decelerating flow observed by MDI/SoHO and GONG \cite{Howe09} leads to a delay in the onset of cycle 24. Moreover, these flux transport dynamo models begin to explore the asymmetry between hemispheres (see 4.3). One important ingredient among others, not
yet accessible to observation, is the number of circulation cells that  describe the meridional circulation (\cite{Jouve07}). A third method, the assimilation method, is also used to better understand what processes guide  the predictions (\cite{Kitiashvili08}).  The three methods  agree rather well but become predictable only when the minimum is reached (Figure \ref{fig:cycle} right).

The 3D simulations of the convective zone including rotation and magnetic field reproduce now reasonably well the latitudinal differential rotation and do not exclude that the circulation patterns could be complex (\cite{Brun04}; \cite{Miesch08}). These 3D simulations are very promising (see Figure \ref{fig:MF}) to understand the interplay between the different processes. This is an opportunity for the community to have two different numerical approaches, one using the Anelastic Spherical Harmonics (ASH) code, the other using the cartesian Pencil code (see Brun and Tavako papers, this issue). The tachocline is probably the seed of the large scale dynamo. The ASH simulations predict that the temperature anisotropy between pole and equator is of about 10 degrees and  is due to the baroclinic effect. Such order of magnitude  seems to be confirmed by HINODE observations.  The multicells obtained in the CZ are favored to understand the cycle lengths of the solar analogs.  

\subsubsection{The deep field and the radiative zones} 
\begin{figure}
\begin{center}
\includegraphics[width=7.cm]{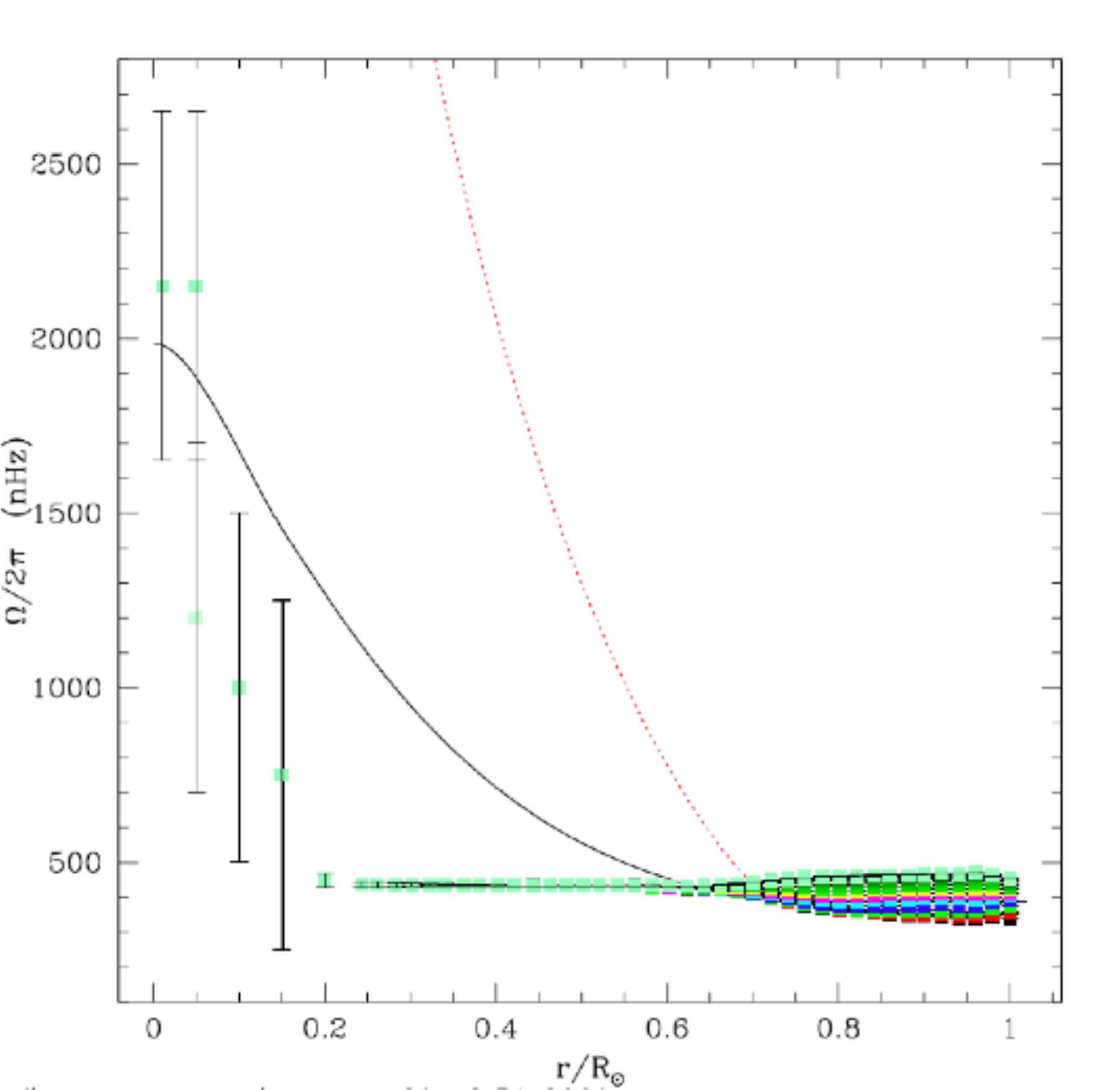}
\caption{Solar internal rotation profile comparison between different models using different initial rotation rate, A: extremely slow (continuous line), B: 20km/s (dotted line) when it arrives on ZAMS, and the profile deduced from the acoustic mode splittings  and potential observed gravity modes. Deduced from Turck-Chi\`eze et al. (2009a).}
\label{fig:sismo}
\end{center}
\end{figure}

Understanding the 11 year solar dynamo is an important issue, but probably not the end of the story of stellar activity. Another fundamental question is the existence of fossil field in deep radiative zone of solar-like stars, knowing that the bulk of the stellar mass (95-98\%) is contained in this part of the stars. The difficulty is due to the absence of any direct known signature of such a field, so rather few works have been dedicated to this fundamental subject during the last decade (\cite{Braithwaite04}; \cite{Zahn07}). Once again, the Sun is the best case where we may find some signature and helioseismology can help us to progress on such subject. An indirect evidence of deep dynamical motions is the solar rotation internal profile, which is extracted from acoustic modes for $\rm R \ge 0.2 R_\odot$  and from gravity modes below this limit. 
Figure \ref{fig:sismo} illustrates what the Sun seems to tell us, compared to solar models that take into account the  transport of momentum and chemicals by rotation (\cite{Turck09}). This paper shows also how different are the meridional circulation in comparison with the convective zone circulations.

Two regions are particularly interesting to comment: the tachocline and the core. 
\begin{figure}
\begin{center}
\includegraphics[width=6.cm]{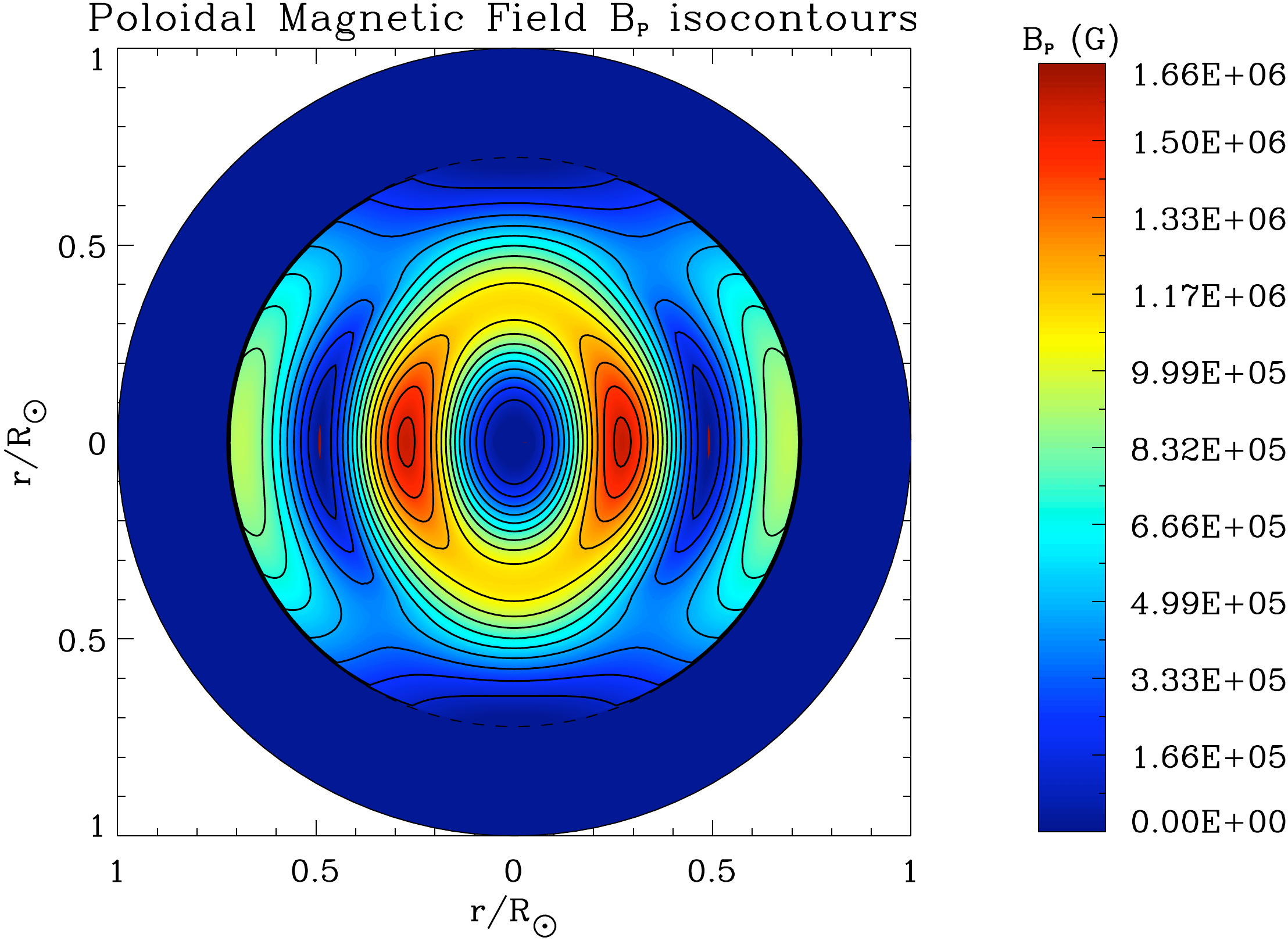}
\includegraphics[width=6.cm]{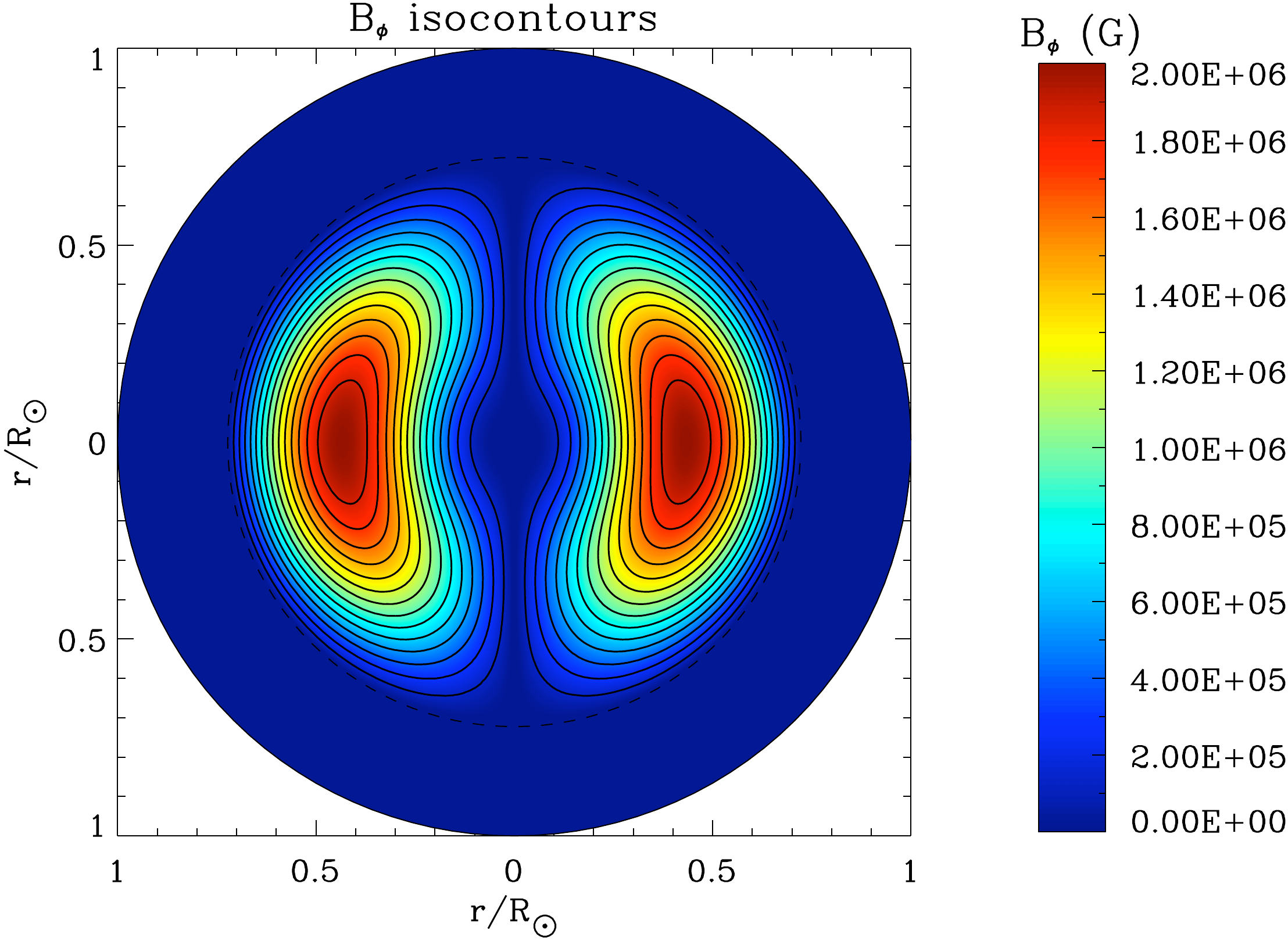}
\caption{Isocontours of the poloidal magnetic field $B_{P(r, \theta)}$ (in G) in meridional cut.
Isocontours of the azimuthal magnetic field $B_{\phi(r, \theta)}$ (in G) in meridional cut for a young Sun. From Duez, Mathis \& Turck-Chi\`eze (2009).}
\label{fig:Du}
\end{center}
\end{figure}

- The tachocline corresponds to the transition layers located between the deep region where the transport of energy is due to radiation and the external region where convection dominates. When photons reach these layers, the interactions with iron, oxygen...are so important that they inhibit the transport of energy, the convection plays the dominant role and partly due to this effect, this region is dominated by shear horizontal turbulence.  Moreover these specific layers probably maintain and restore the convective toroidal magnetic field. This results in a suppression of the differential rotation observed in the whole convective zone. As shown on Figure \ref{fig:sismo}, it is now well established that the radiative zone rotates as a solid body at least down to the limit of the  energy generation core (\cite{Eff-Darwich08}). 

- The solar core has not yet received so much attention due to the absence of clear dynamical information up to recently. But the GOLF instrument aboard the SoHO satellite appears more and more promising to detect several individual gravity modes, the reason is given in \cite{Turck08}. The present searches (\cite{Turck04b}; \cite{Garcia07}; Mathur et al., 2007, 2008) favor a rotation in the core between 2 to 5 times the rotation of the rest of the radiative zone. This tendency might be confirmed before the end of the SoHO observations by detecting individual splitting with the signal of more than 13 years of data. In fact, individual components of dipolar modes become  reachable at low frequency, these modes are strongly dependent on the central core rotation. 
Figure \ref{fig:sismo} shows that the induced rotation profile cannot be explained only by the effect of the transport of momentum and chemicals by rotation, so the natural first candidate to flatten the profile is the fossil magnetic field. Gough \& McIntyre suggested that the uniformity of the rotation profile observed in
the bulk of the radiative zone is due to the presence of a large-scale primordial magnetic
field confined below the tachocline by flows originating from  the convection
zone (\cite{Gough09}). This idea has been recently numerically simulated by \cite{Garaud08}. They emphasize the key role of
flows downwelling from the convection zone in confining the assumed internal field and the quenching of the large-scale differential rotation below the tachocline, including
in the polar regions, as seen by helioseismology. A new approach illustrated by Figure  \ref{fig:Du}
is to determine a stable magnetic configuration (a mixture of poloidal and toroidal fields) in order to follow its diffusivity and its transport of momentum along the long lifetime of the solar-like stars (\cite{Duez09}). This work will lead to a complete and dynamical view of the solar interior.

{\it The interaction between the fossil field and the dynamo field is not yet explored and certainly much work needs to be dedicated to the radiative zone. }

\begin{figure}
\begin{center}
\vspace{-1.5cm}
\includegraphics[width=12.cm]{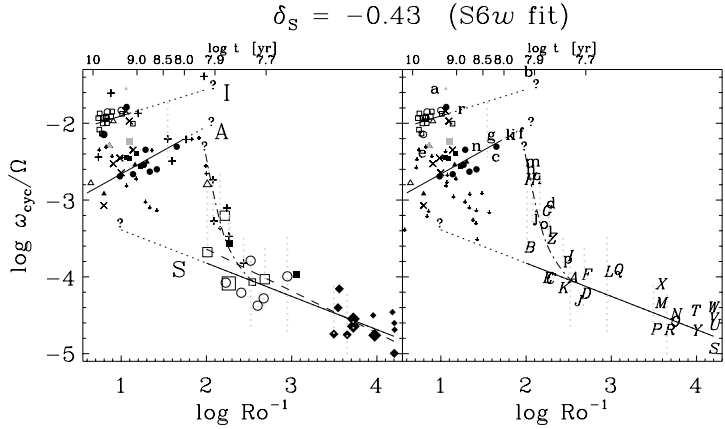}
\vspace{-7.5cm}
\caption{Relationship between cyclicity, rotation and the Rossby number (R$_{\rm o}$ = P$\rm_{rot}/\tau_c$) for an extended "echantillon" of stars. From Saar and Brandenburg (1999).}
\label{fig:stars}
\end{center}
\end{figure}

\subsection{The young stars and the solar analogs} 
Many stars show magnetic activities and more and more observational methods are available. Chandra and XMM allow observations of Alpha Centauri which are twins of the Sun and of a lot of other stars for which coronal observations are possible. Several other indicators like CaII lines (variation of 2\% in the solar case), Doppler imagers, spectropolarimetry with ESPADON, NARVAL and SEMPOL give complementary information like for example the presence of spots at the pole of young stars or some potential geometry of the external magnetic field. At Mount Wilson Observatory, on 100 stars followed along 25 years, 60 exhibit cyclicity, 25 are variables and 15 do not show any activity (Baliunas, 1995). 

The study of \cite{Saar99}, summarized by Figure \ref{fig:stars}, shows that stars are not strictly segregated onto one or the other branch by activity level. The high-cycle branch is primarily composed of inactive stars. The extended dataset suggests that after 1 Gyr, stars can have cycles on one or both branches,
though among older stars, those with higher (lower) mass tend to have their primary P$_{cyc}$ on the lower (upper) cycle branch. The solar  Gleissberg cycle agrees with this scenario, suggesting that long
term activity trends,  in many stars, may be segments of long cycles not yet resolved
by the data. Most of the very active stars ($P_{rot} \ge$ 3 days) appear to occupy a new third branch, with ${\omega_{cyc}/\Omega}$ proportional to $\rm R_o^{0.4}$. Many RS CVn variables lie in a transition region between the two most active branches. These authors  predict P$_{cyc}$ for three groups: stars with long-term Ca II HK trends, stars in young open clusters, and stars that may be in Maunder-like magnetic minima.

Observations of young stars have grown sharply, they are generally rapidly rotating stars and many possess substantial magnetic activity and strong axisymmetric magnetic fields. These stars are particularly interesting to follow in order to understand the young stage of the Sun, and also to better understand the impact of young stars on planets. 
\begin{figure}
\begin{center}
\includegraphics[width=7cm]{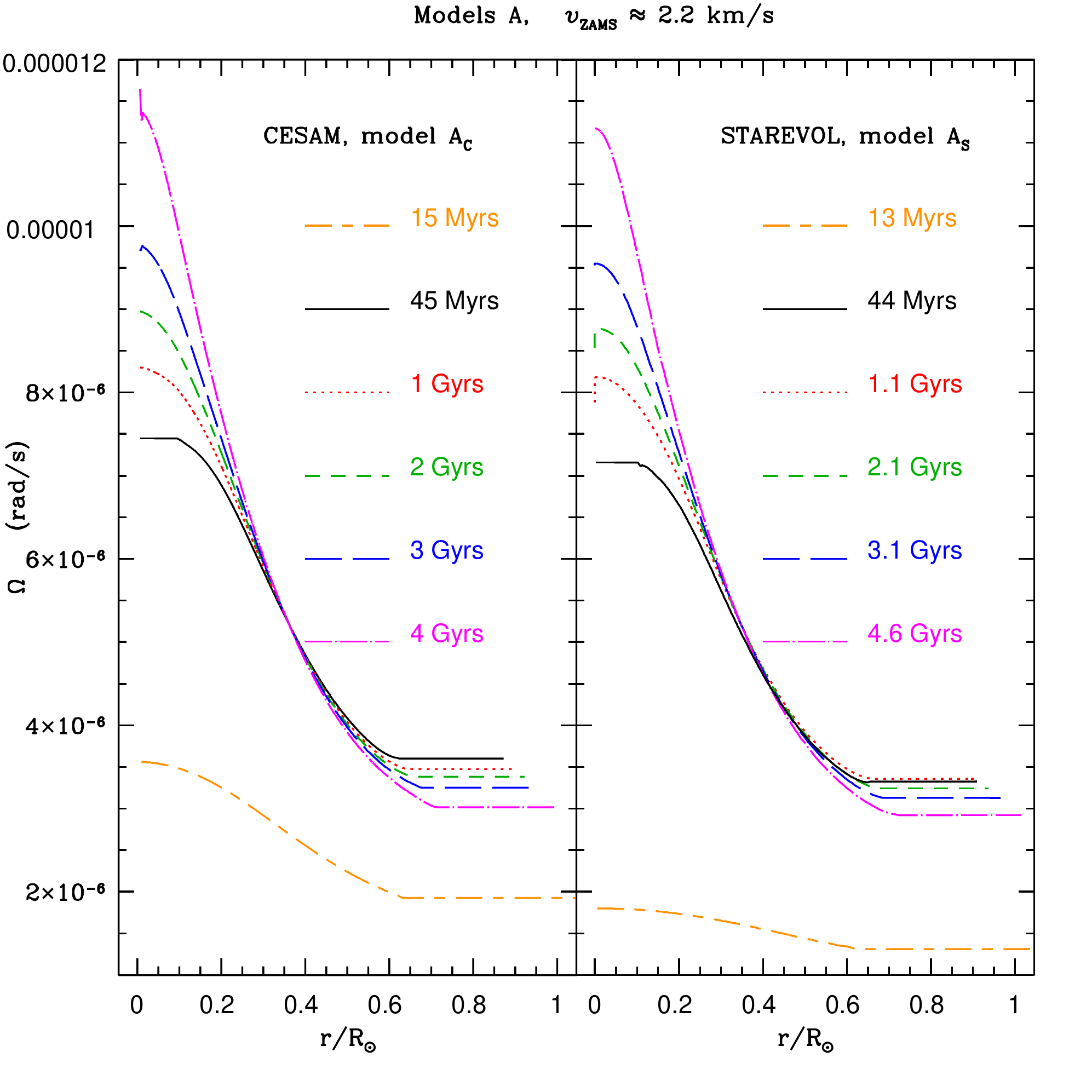}%
\includegraphics[width=7cm]{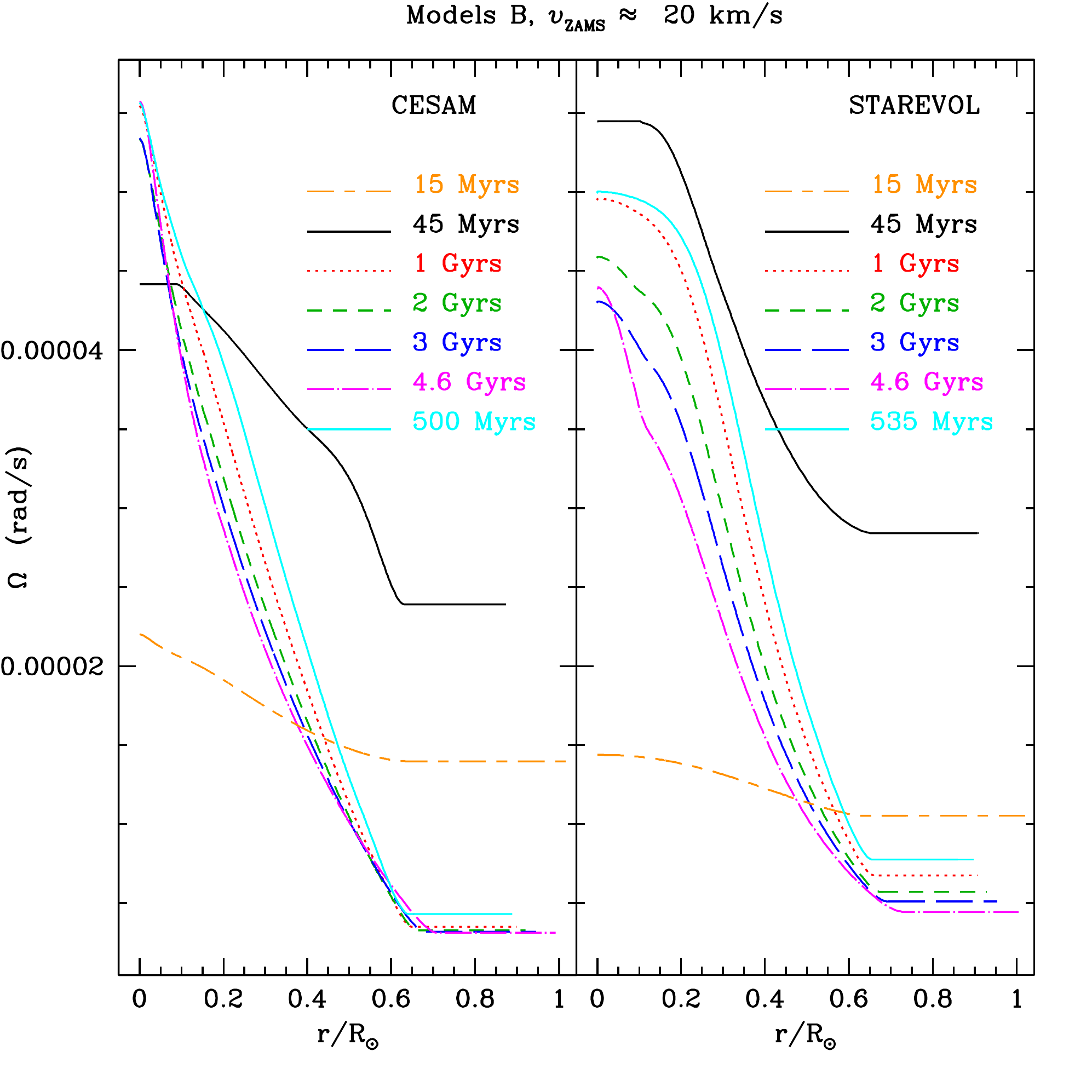}
\caption{Evolution of the angular velocity profiles from the contraction phase of a star of 1 M$_\odot$ at six different ages
obtained by two 1D evolution codes: CESAM and STAREVOL, for an extremely slow initial rotation rate and for a moderate case where the star arrives on the ZAMS with a surface rotation of 20 km/s and a sudden magnetic braking. The age of 45 Myr corresponds to the arrival on the main sequence. From Turck-Chi\`eze et al. (2009a).}
\label{fig:evolomega}
\end{center}
\end{figure}

Figure \ref{fig:evolomega} shows how the angular velocity profiles evolve with time for young solar-like star, potential young Sun. One notes that the radial contrast increases during the PMS phase even if the initial rotation is not so large due to the contraction phase. The advective term in the transport of momentum, connected to the meridional circulation, can significantly increase this contrast in the radiative zone, in opposite to the role of the magnetic braking which is generally invoked to explain the strong reduction of the superficial rotation during the main sequence (\cite{Turck09}). But in the two cases, the radial differential rotation during the main sequence is increased by no more than 20\% or 50\% of the initial rotation rate achieved at the end of the PMS phase.  

In parallel, simulations, with the 3D hydrodynamics ASH code,  have shown that the young rapidly rotating stars may 
know intermittent convective phases (\cite{Ballot07}). The team of  {\it Sun in time} has studied  about 13 stellar analogs of the Sun of about 1 M$_\odot$ but with rather different ages and rotation rates varying from 1 to 35 days. They exhibit very different types of dynamo, radial and often saturated dynamos (\cite{Gudel07}). Simulations of dynamo action in rapidly rotating suns with the  ASH code  explore the complex coupling between rotation, convection and magnetism. These simulations show  that substantial organized global-scale magnetic fields are achieved by dynamo action in these systems. Striking wreathes of magnetism are built in the midst of the convection zone, coexisting with the turbulent convection and in the absence of tachocline.  Some dynamos achieved in these rapidly rotating stars build persistent global-scale fields which maintain their amplitude and polarity for thousands of days. For five times the solar rate, the dynamo can undergo cycles of activity, with fields varying in strength and even changing polarity. As the magnetic fields wax and wane in strength, the primary response in the convective flows involves the axisymmetric differential rotation, which begins to vary on similar time scales. Bands of relatively fast and slow fluid propagate toward the poles on time scales of roughly 500 days. In the Sun, similar patterns are observed in the poleward branch of the torsional oscillations, and these may represent a response to poleward propagating magnetic field deep below the solar surface (\cite{Brown10}).

{\it The history of the magnetic field from the formation of the star to the solar present age is a very missing stone which must require more attention from both the observational side and the  theoretical, modelling and simulation sides.}

\section{Stars and Planets interaction (s)}
\begin{figure}
\begin{center}
\includegraphics[width=9.cm]{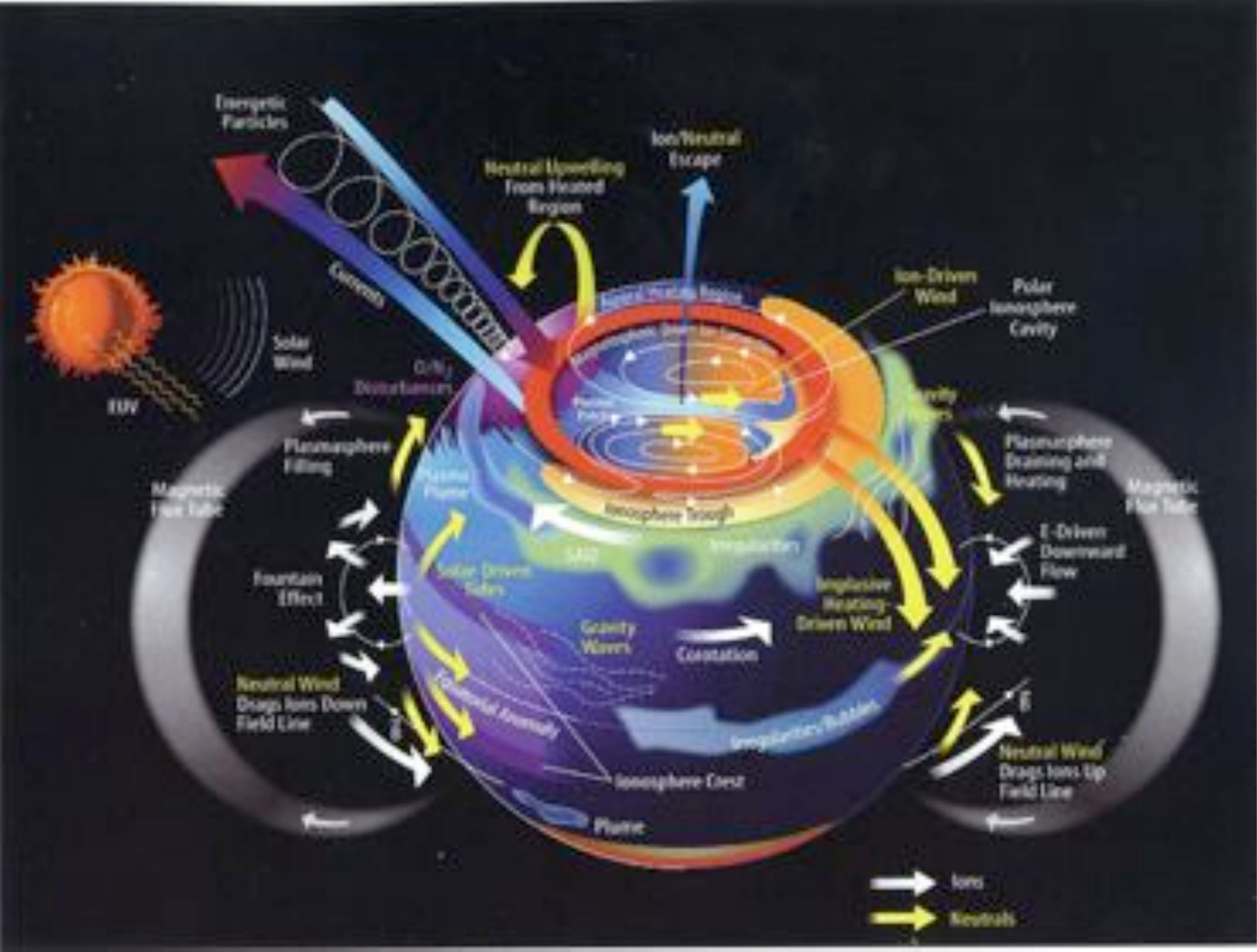}
\caption{This figure illustrates the complexity of the  phenomena which govern the Sun-Earth relationship. Courtesy of Joseph Grebowsky, NASA GSFC.}
\label{fig:ESinteraction}
\end{center}
\end{figure}

The interest to study Sun and solar-like star activity is amplified by the interest to describe the complex interaction between the Sun (Stars) and the Earth (Planets). Figure \ref{fig:ESinteraction} summarizes the complexity of these interactions. It  also shows that the physical processes which describe the stellar and external stellar dynamics and the processes which need to be understood in the Earth atmosphere are inspired by the same physics. This field is rapidly growing thanks to the capability to detect planets by Doppler velocity techniques. Accordingly, we may hope that a common community will appear in the next decade to understand  the real impact of the Sun (Stars) on the Earth (Planets) environment and climate.  Precise space luminosity variations with MOST, COROT, KEPLER and in the future PLATO will contribute to the development of this field. So a new interest has been developed this last decade to understand the conditions of life on planets.

\subsection{Interaction between young stars and planets}
In this context, young stars are again under the spotlight.  Indeed, it is important to observe analogous stars of the Sun to tentatively find the kind of stars which can favor life on planets. A sequence formed by EK Dra (100Myr), $\pi^1$ UMA (300 Myr), $\kappa^{-1}$ Cet (650 Myr) and the Sun delivers many informations on the interaction between young stars and their planets (\cite{ Kaltenegger09}). The young stars are rotating faster than the Sun and a sequence of rotation can be considered (see Ribas talk). For the youngest, X rays and XUV are 100 to 1000 more intense than the solar case, the visible light is reduced by about 30\%, FUV and UV are 5 to 60 times the present value, the solar wind 10 to 1000 the present solar one and the flares more frequent too (about 10 times per day). So clearly the young Sun  or young solar-like stars interact with their planets rather differently and rather strongly. 

As a natural consequence Lyman $\alpha$, FUV and UV emissions produce photo dissociation reactions of CO$_2$, H$_2$O, CH$_4$, NH$_3$. X rays, EUV and Lyman $\alpha$ emissions heat, expand and photoionize the exosphere. The intense stellar wind carries away more atmospheric particles and erodes the atmosphere of the planet.  If now one looks for the type of stars that can support habitable planets, it is noticed that due to biological damage by UV, it seems today that they are restrained to F and G stars. It is then interesting to look to twins of the Sun. Among others, $\kappa$ Ceti appears  particularly similar to the Sun (\cite{Porto06}).

{\it There is evidence that this science will largely progress with the KEPLER satellite these coming years. Building an history of the Sun-Earth connection along the evolution of the star  will help to build an history of habitable planets.}

\subsection{The Sun-Earth relationship and the Space Weather}
This discipline is rather young. It is composed of two parts. The first one,  the quasi instantaneous relationship (at the level of seconds, hours or days) also called space weather, is now well organized since the launch of SoHO. The other one, at the level of months, years, decades or centuries, is still in its infancy and will gain to be developed.  The frontier between these two aspects is not so clear and one can imagine that it will further reduce in the future (see the last section).
\subsubsection{Space weather, X, $\gamma$ and radio emissions}
Space weather regroups the conditions of the Sun and in the solar wind, magnetosphere, ionosphere and thermosphere that can influence the performance and reliability of space-borne and ground-based technological systems and can endanger human life or health.
The most energetic phenomena of solar activity are flares and coronal mass ejections.
Flares are large explosions that occur over the solar atmosphere and may last from a few seconds
to hours. A solar flare is caused by a sudden, and yet unpredicted, energy release high above
the magnetic loops. This magnetic energy is then transformed into particle acceleration (through their kinetic energy) and heat
of the surrounding atmosphere. Both the energetic particles and the hot gas produce emission
throughout the whole electromagnetic spectrum, from the very energetic gamma-rays all the way
to long radio waves. From the observation of the emission produced during flares it is possible
to infer the energetic particles spectrum and thus get a clue on the acceleration mechanism that
produces these particles. The recent findings of flare observations, in the gamma-rays domain by the RHESSI
satellite and at high radio frequencies by the Solar Submillimeter Telescope, are now analyzed. The observations of the geometry of the localized flux emergence need space and time resolution,
the understanding of the underlying physics is now largely worked out (See Valio this issue; \cite{Wang}).

\subsection{The solar impact on the terrestrial climate}
This subject is extremely controversial and difficult due to the required accuracy, the use of proxies and the competition between the different influences.

- direct influence:  

The solar irradiance contributes directly to the Earth temperature, consequently its variability is of prime importance. The Sun is not far from a black body so its total irradiance variability is dominated by the visible range variability. Such quantity is now followed since 3 decades, but with different intruments and with problems of ageing and calibration. The variability due to plages and rotation is well measured (up to 3-4 W/m$^2$) but the comparison between  minima is still in debate but remains smaller than 0.5 W/m$^2$ (\cite{Scafetta09}; \cite{Krivova09}) compared to 1361- 1367 W/$^2$. Consequently the impact of the irradiance variability on the Earth climate is considered today as rather small, unless the variability   was amplified during maxima by other phenomena. 

The extrapolation towards the past is not  easy  and has been readjusted several times, leading each time to lower variation. From the $^{14}$C measurements obtained in different sites (see Miyahara talk) showing clearly anti-correlation  with the solar activity due to the solar dynamo, it has been shown that this dynamo persists even during long minima.  Moreover  the last millenium can be reasonably studied in using the open flux, deduced from the geomagnetic aa-index. After scaling on the last decades, it allows to extrapolate our knowledge of the total irradiance down to the Maunder minimum (at least) leading to a change of no more than 1.3 W/m$^2$ \cite{Krivova07}.

{\it Could this irradiance change enough to justify the observed decrease of the mean Earth temperature during the Maunder minimum?}

- indirect dynamical influence: 

It is more and more believed that the Solar-Earth relationship on long timescale (decades or centuries)  is largely more complex than previously thought (\cite{Haigh05}). It appears insufficient  to consider only an energetic relationship. It is well known that the X, EUV, UV parts of the solar spectrum vary much more strongly than the total irradiance, but their impact cannot be seek in through the energetic changes, which are this time extremely small. So the magnetic interaction, the solar wind and the dynamics of the heliosphere are more and more studied to analyze the impact on the ionosphere and one the chemistry of the atmosphere (through solar wind or subsequent cosmic rays variation). Such variability may contribute to some indirect and dynamical effects. Today, this field of interaction begins to be explored and justifies to look to Earth latitudinal and seasonal effects. This supposes having robust and complete climatic models, like LMDz-Reprobus which includes full representations of dynamical, radiative, and chemical processes in the atmosphere and their interactions, especially feedbacks of the chemical tendencies on the dynamics : in particular, ozone is strongly affected by dynamics and transport (\cite{Jourdain08}).

\section{Open questions and Perspective}
This review has tried to summarize the present situation of this thematic. 

From this meeting, a lot of questions emerge naturally. 
\begin{itemize}
\item 
  Could we establish a proper relationship between luminosity variation and solar radius, can we measure solar surface deformation and deduce subsurface magnetic field variability and the order of magnitude of the deep field if it exists ?
\item
Is the solar dynamo sufficiently understood today ? Could many dynamos exist in the Sun ? one dynamo (global scale), two dynamos (shallow + CZ global scale) or three dynamos (with also a core dynamo) ?
\item
Is the magnetic diffusivity at the origin of the de-synchronization between the two hemispheres. Does it contribute to the lifetime of the cycle (\cite{Guerrero09}) ?
\item
Is the radiative zone only a stable zone, or does it contain a fossil field which participates to the variability.  Does it contribute to the greater cycles ($\ge$ 80 years) through convective plums exciting gravity waves ? Is the fossil field that initially creates the toroidal field of the tachocline ?
\item
Do we need to understand the whole dynamics of the Sun to build the bridge between active young stars and Gyr stars in exploring the different types of dynamos ? 
\item
Is the Earth atmosphere more  influenced by the solar dynamics and the particle fluxes than by the irradiance variability ? Is it also true for Star-Planet relationship just after the formation epoch ?
\item
Are the present studies for the existence of life on planets sufficiently mature to orientate the present search on exoplanets and are we ready for the next objectives ?
\end{itemize}

Some of these questions will find an answer soon.
With the  launch of PICARD and SDO in 2010, we hope to establish the absolute solar luminosity which is surprisingly not known today better than 0.4\%. The impact of this uncertainty on a solar model is important, such difference corresponds to 100 Myrs or a boron neutrino flux modified by 5\%.
We hope also to progress on the shape of the Sun which stays one of the best indicators with the solar wind that put interesting constraints on the strength and geometry of the internal field (s).  The dynamics of the CZ will be largely improved.

In parallel to the observations, we need to cumulate numerical exercises and confront them to the observational facts. With the present capability of computers, we need to develop secular stellar evolution models with transport of momentum and chemicals by dynamical processes in 1 or 2D in parallel to 3D simulations which focus on the interplay between dynamical scales. For the solar-like stars, the 1D dynamical models are totally justified because the stellar structure  is only slightly modified by the presence of internal dynamical phenomena and the dynamical processes evolve on secular scales.
\begin{figure}
\begin{center}
\includegraphics[width=5.5cm]{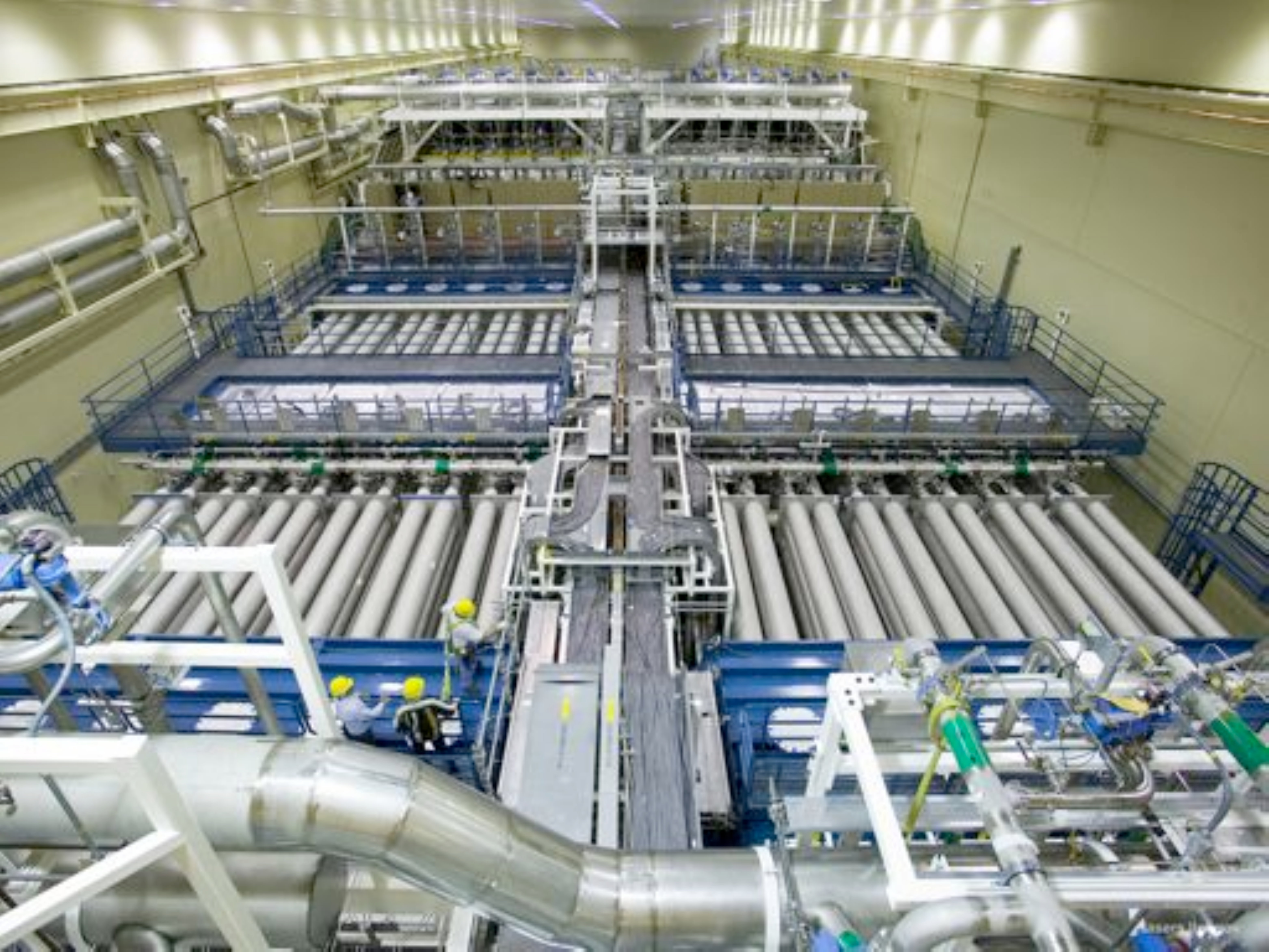}
\includegraphics[width=7.5cm]{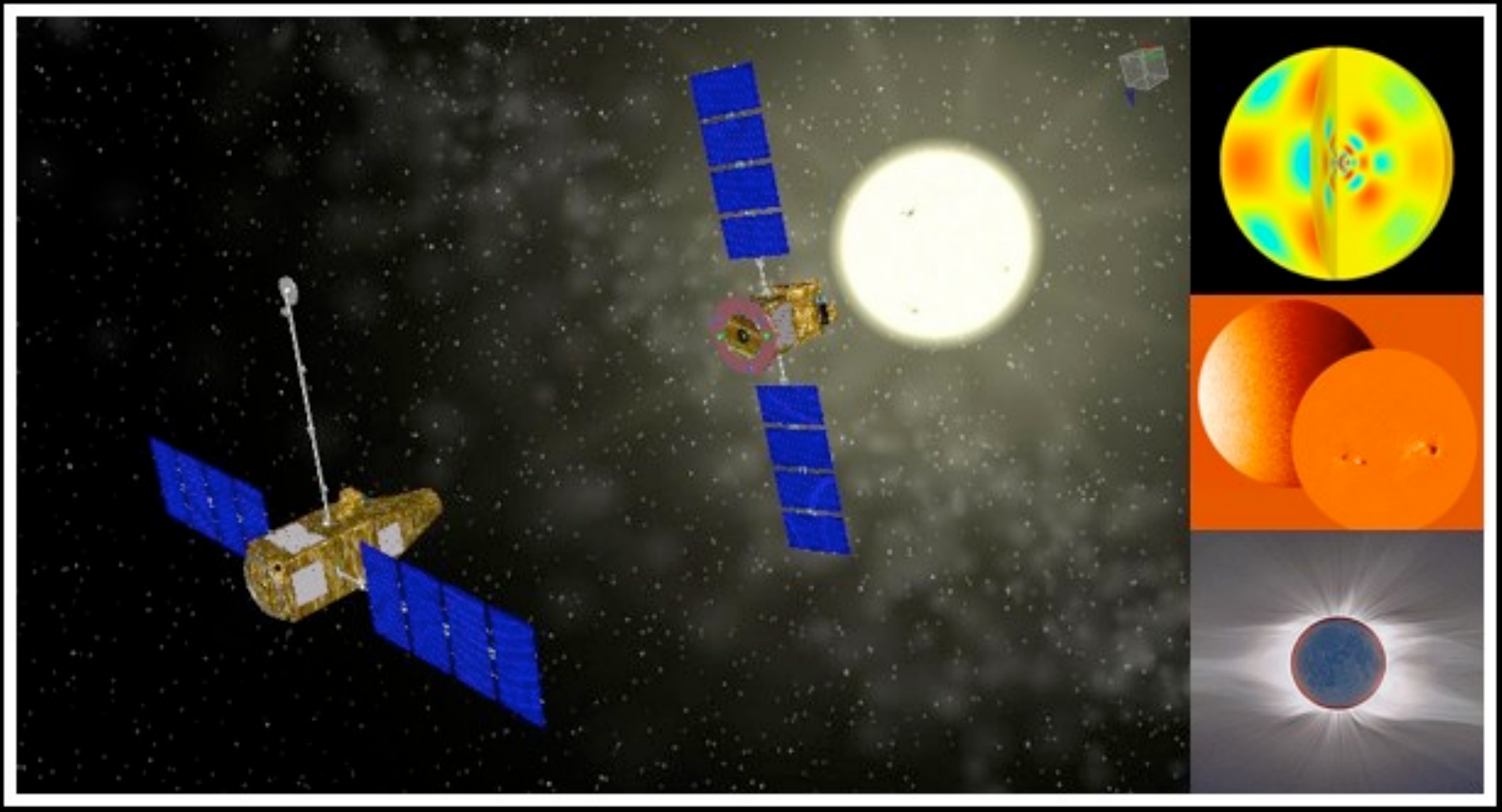}
\caption{a) Inertial fusion study in the NIF hall https://lasers.llnl.gov/. b) The formation flying concept, an excellent perspective for the simultaneous development of Space Weather and Space climate (\cite{Turck09b}).}
\label{fig:projects}
\end{center}
\end{figure}
\vspace{0.5cm}

Let me now express my personal thoughts:
\begin{itemize} 
\item
The standard solar model is no more the right framework for the comparison with observations. It remains nevertheless an excellent basis on which one must integrate step by step the effect of rotation, magnetic field and gravity waves. Until the arrival of the complete Dynamical Solar Model (DSM), the seismic model (SeSM), built to reproduce the sound speed profile (\cite{Couvidat04};  \cite[\turck et al., 2004a]{Turcketal04a};  \cite{Mathur07}),  is the best solar model for structural and observable predictions. 
\item
Secular dynamical models need to be developed (1D-2D) for a good understanding of all the sources of variability (in particular at larger scale than the 22 yrs). Solar  (especially gravity mode detection), young stars, solar analogs observations will  enrich the related physics (topology of the internal magnetic field). After COROT and ESPADON, KEPLER represents an impressive step forward.
Cartesian and Spherical 3D simulations must be pursued, they will benefit from the increased power of the parallel computers to treat the very different scales of the solar (stellar) interiors.
\item
There was a lot of attention on  solar sunspots (a long series) but dynamo exists without them, more focus on the radiative zone and probably on the corona flow is necessary to build the complete magnetic field story.
\item
There is a large need  to build time series of other emissions: X rays, CME (in minima, not related to ARÉ), big flares, UV, for their use with data assimilation methods for the development of Space Weather and Space Climate. One needs to establish clear relation between stars and planets: formation, atmosphere and possible life.
For the present Sun-Earth relationship we begin to have good daily information. But the evolution of the climatic relationship is not yet sufficiently established to give quantitative values of the influence of competitive processes: the Sun must certainly not be studied only as acting as a global energetic external input but one needs to build the seasonal and latitudinal influences of a lot of varying indicators. This requires the use of detailed 3D dynamical and chemistry terrestrial atmospheric models which are emerging nowadays.
\item
The facination for the Sun and Stars has not decreased due to fundamental and societal reasons (magnetic fusion, inertial fusion, space weather and space climate, life on planets). Big laser installations  begin to produce stellar plasmas in laboratory to study micro and macro properties (Figure \ref{fig:projects} left).
\item
We need to pursue  continuous observation of the Sun. Rather than a lot of small satellites, it would be extremely exciting to organize ourselves to build a large world class mission with two Herschel-Planck platforms organized in formation flying (Figure \ref{fig:projects} right) at L1 point. Such approach allows a permanent eclipse with an insight on the low corona from where the solar flux emerges. But  the advantage of this concept is also to carry all the instruments we need in order to continuously look to the Sun from the core to the corona at different spatial and temporal scales. Such mission must be more ambitious than  the project proposed in the framework of  Cosmic Vision (\cite{Turck09b}). It is the only concept that contributes to the simultaneous development of Space Weather and Space climate. PROBA3 is the first step in this direction.
\item
The investigation of young stars is very important. It must be amplified together with properties of exoplanets. The whole stellar magnetic field story is an important objective for the next decade. 
\end{itemize} 
\acknowledgement{I would like to thank the organizers who have offered to all the participants, a program that shows the whole chain of  processes. The result has been a very enjoying meeting. I would like to add personal warm thanks to all the participants who are the main actors of this wonderful week in Rio de Janeiro; they have inspired  this present complementary review. I acknowledge the Wilcox Solar Observatory (WSO) of Stanford University directed by Prof. P.H. Scherrer for use of their polar field observations.  A lot of results are due to space experiments from ESA, NASA and JAXA, we are grateful to these agencies and to national agencies like CNES as they  not only contribute to the realization of  instruments but they also accompany us  in the scientific return. }

\end{document}